\pdfoutput=1
\documentclass[graybox]{svmult}

\usepackage[utf8]{inputenc}
\usepackage{url}

\usepackage{amsmath,array}
\usepackage{amssymb}
\usepackage{subfig}
\usepackage[section]{placeins}

\usepackage{todonotes}

\usepackage{type1cm}        
\usepackage{makeidx}         
\usepackage{graphicx}        
\usepackage{multicol}        
\usepackage[bottom]{footmisc}

\usepackage{listings}
\usepackage[scaled=0.9,l]{zlmtt} 

\usepackage{newtxtext}       %
\usepackage{newtxmath}       

\usepackage{tikz}
\usetikzlibrary{calc}
\usetikzlibrary{math}
\usepackage{pgfplots}
\pgfplotsset{compat=1.14}
\usepackage{adjustbox}

\usepackage{algorithm}
\usepackage{algorithmicx}
\usepackage[noend]{algpseudocode}
\usepackage{tabu,booktabs}
\usepackage{tabularx}
\usepackage[detect-weight,binary-units]{siunitx}

\DeclareSIUnit\dof{DOF}
\DeclareSIUnit\flop{FLOP}

\newcolumntype{Y}{%
  S[
  table-format=1.2e1,
  table-auto-round,
  scientific-notation=true,
  ]}

\newcolumntype{F}{%
  S[
  table-format=2.1,
  round-mode=figures,
  round-precision=3,
  ]}

\newcolumntype{G}{%
  S[
  table-format=3,
  round-mode=figures,
  round-precision=3,
  ]}

\definecolor{dunedarkblue}{rgb}{0.22, 0.29, 0.49}
\definecolor{dunemediumblue}{rgb}{0.35, 0.45, 0.71}
\definecolor{dunelightblue}{rgb}{0.61, 0.68, 0.85}
\definecolor{duneorange}{rgb}{0.91, 0.56, 0.25}

\definecolor{darkgreen}{rgb}{0., 0.5, 0.0}
\definecolor{darkblue}{rgb}{0., 0., 0.5}
\definecolor{darkred}{rgb}{0.5, 0., 0.}
\definecolor{darkorange}{rgb}{1., 0.35, 0.0}

\definecolor{niceblue}{rgb}{0.122,0.396,0.651}   
\definecolor{niceorange}{RGB}{255,205,86}        
\definecolor{nicered}{RGB}{220,20,60}                      
\definecolor{niceteal}{HTML}{00A9AB}
\definecolor{niceviolet}{HTML}{820080}

\definecolor{niceblueLight}{HTML}{91CAFB}
\definecolor{niceblueVeryLight}{HTML}{DDEFFF}

\colorlet{block1}[rgb]{darkred}
\colorlet{block2}[rgb]{darkred>wheel,1,4}
\colorlet{block3}[rgb]{darkred>wheel,2,4}
\colorlet{block4}[rgb]{darkred>wheel,3,4}

\usepackage{xspace}
\newcommand{\exadune}{\textsc{Exa-Dune}\xspace}
\newcommand{\dune}{\textsc{Dune}\xspace}
\newcommand{\dunemodule}[1]{\textsc{Dune-#1}\xspace}
\newcommand{\istl}{\dunemodule{ISTL}\xspace}
\newcommand{\pdelab}{\dunemodule{PDELab}\xspace}

\DeclareMathOperator{\diam}{diam}

\usepackage[final]{changes}

\makeindex             

\begin{document}

\title*{\exadune{} --- Flexible PDE Solvers, Numerical Methods and Applications}

\author{Peter Bastian\inst{1}
\and Mirco Altenbernd\inst{3}
\and Nils-Arne Dreier\inst{2}
\and Christian Engwer\inst{2}
\and Jorrit Fahlke\inst{2}
\and René Fritze\inst{2}
\and Markus Geveler\inst{4}
\and Dominik Göddeke\inst{3}
\and Oleg Iliev\inst{5}
\and Olaf Ippisch\inst{6}
\and Jan Mohring\inst{5}
\and Steffen Müthing\inst{1}
\and Mario Ohlberger\inst{2}
\and Dirk Ribbrock\inst{4}
\and Nikolay Shegunov\inst{5}
\and Stefan Turek\inst{4}
}

\authorrunning{Peter Bastian et. al.}

\institute{%
\inst{1} Interdisciplinary Center for Scientific Computing, Heidelberg University,
Im Neuenheimer Feld 368, D-69120 Heidelberg,
\inst{2} Institute for Analysis and Numerics, University of Münster,
Orleans-Ring 10, D-48149 Münster,
\inst{3} Institute for Applied Analysis and Numerical Simulation, University of Stuttgart,
Allmandring 5b, D-70569 Stuttgart,
\inst{4} Inst. f. Applied Mathematics, TU Dortmund,
Vogelpothsweg 87, D-44227 Dortmund,
\inst{5}Fraunhofer Institute for Industrial Mathematics ITWM
Fraunhofer-Platz 1, D-67663 Kaiserslauter,
\inst{6}Institut für Mathematik, TU Clausthal-Zellerfeld,
Erzstr. 1, D-38678 Clausthal-Zellerfeld.
}

\maketitle

\abstract{In the \exadune{} project we have developed, implemented and optimised
numerical algorithms and software for the scalable solution of partial
differential equations (PDEs) on future exascale systems exhibiting
a heterogeneous massively parallel architecture.
In order to cope with the increased probability of hardware failures, one
aim of the project was to add flexible,
application-oriented resilience capabilities into the framework.
Continuous improvement of the
underlying hardware-oriented numerical methods have included GPU-based sparse approximate inverses,
matrix-free sum-factorisation for high-order discontinuous Galerkin
discretisations as well as partially matrix-free preconditioners.
On top of that, additional scalability is facilitated by exploiting
massive coarse grained parallelism offered by multiscale and uncertainty
quantification methods where we have focused on the adaptive choice of the
coarse/fine scale and the overlap region as well as the combination of
local reduced basis multiscale methods and the multilevel Monte-Carlo
algorithm. Finally, some of the concepts are applied in a land-surface model including subsurface flow 
and surface runoff.
}

\section{Introduction}


In the \exadune{} project we extend the Distributed
and Unified Numerics Environment (DUNE)\footnote{\url{http://www.dune-project.org/}} ~\cite{goeddeke_mini_para:dunegridpaperI,goeddeke_mini_para:dunegridpaperII} by
hardware-oriented numerical methods and hardware-aware implementation techniques developed
in the \replaced{(now) FEAT3}{FEAST}\footnote{\url{http://feast.tu-dortmund.de/}}~\cite{goeddeke_mini_para:feast}
project to provide an exascale-ready software framework for the numerical
solution of a large variety of partial differential equation (PDE) systems with state-of-the-art numerical methods including higher-order
discretisation schemes, multi-level iterative solvers, unstructured and locally-refined meshes, multiscale
methods and uncertainty quantification, while achieving close-to-peak performance and exploiting
the underlying hardware.

In the first funding period we concentrated on the node-level performance as
the framework and in particular its algebraic multigrid solver already show very
good scalability in MPI-only mode as documented by the inclusion of DUNE's solver library in the
High-Q-Club,
the codes scaling to the full machine in Jülich at the time, with close to half a million cores.
Improving the node-level performance in light of future exa-scale hardware
involved multithreading (``MPI+X'') and in particular exploiting SIMD parallelism (vector extensions
of modern CPUs and accelerator architectures). These aspects were addressed within the finite element assembly
and iterative solution phases.
Matrix-free methods evaluate the discrete operator without storing a matrix, as the name implies,
and promise to be able to achieve a substantial fraction of peak performance.
Matrix-based approaches on the other hand are limited by
memory bandwidth (at least) in the solution phase and thus typically exhibit only a small fraction of the peak (GFLOP/s)
performance of a node, but decades of research have lead to robust and efficient (in terms
of number of iterations) iterative linear solvers for practically relevant systems.
Importantly, a consideration of matrix-free and matrix-based methods needs to take the order of the
method into account. For low-order methods it is imperative that a matrix entry can be
recomputed in less time than it takes to read it from memory\added{,
to counteract the memory wall problem}.
This requires to exploit the problem
structure as much as possible, i.e., to rely on constant coefficients, (locally) regular mesh structure and
linear element transformations
\cite{gmeiner2016quantitative,kohl2018hyteg}.
In these cases it is even possible to apply stencil type techniques,
like developed in the EXA-STENCIL project \cite{exastencils}.
On the other hand, for high-order methods with tensor-product
structure the complexity of matrix-free operator evaluation can be much less than that of matrix-vector
multiplication, meaning that less floating-point operations have to be performed which at the same time
can be executed at a higher rate due to reduced memory pressure and better suitability for vectorization
\cite{ORSZAG198070,Buis:1996,Kronbichler2012135}.
This makes high-order methods extremely attractive for exa-scale machines \cite{2017arXiv171110885M,PIATKOWSKI2018220}.

In the second funding phase we have mostly concentrated on the following aspects:
\begin{enumerate}
\item \textbf{Asynchronicity and fault tolerance:} High-level C++ abstractions form the basis of transparent error handling
using exceptions in a parallel environment, fault-tolerant multigrid solvers as well as communication hiding Krylov methods.
\item \textbf{Hardware-aware solvers for PDEs:}
We investigated matrix-based sparse-approximate inverse preconditioners including novel machine-learning
approaches, vectorization through multiple right-hand sides as well as matrix-free high-order Discontinous Galerkin (DG) methods and partially
matrix-free robust preconditioners based on algebraic multigrid (AMG).
\item \textbf{Multiscale (MS) and uncertainty quantification (UQ) methods:} These methods provide an additional layer
of embarrassingly parallel tasks on top of the efficiently parallelized forward solvers. A challenge here is load balancing of the asynchronous
tasks which has been investigated in the context of the localized reduced basis multiscale method and multilevel
Monte Carlo methods.
\item \textbf{Applications:} We have considered large-scale water transport in the subsurface coupled
to surface flow as an application where the discretization and solver components can be applied.
\end{enumerate}

In the community, there is broad consensus on the assumptions about exascale systems
that did not change much during the course of this six year project.
A report by the Exascale Mathematics Working Group to the U.S. Department of
Energy's Advanced Scientific Computing Research Program~\cite{Dongarra:2014:AMR}
summarises these challenges as follows, in line with \cite{Keyes:2011:ETW} and more recently the Exascale Computing Project\footnote{\url{https://www.exascaleproject.org/}}: \\
(i) The anticipated power envelope of 20\,MW implies strong limitations on the amount and organisation of the hardware components, an even stronger necessity to fully exploit them, and eventually even  power-awareness in algorithms and software.
(ii) The main performance difference from peta- to exascale will be through a 100--1000 fold increase in parallelism at the node level, leading to extreme levels of concurrency and increasing heterogeneity through specialised accelerator cores and wide vector instructions.
(iii) The amount of memory per \lq core\rq\ and the memory and interconnect bandwidth / latency will only increase at a much smaller rate, hence increasing the demand for lower memory footprints and higher data locality.
(iv) Finally, hardware failures, and thus the mean-time-between-failure (MTBF), \replaced{were}{are} expected to increase proportionally (or worse)
corresponding to the increasing number of components.
\added{Recent studies have indeed confirmed this expectation~\cite{Gupta:2017:FIL}, although not at the projected rate.}
First exascale systems are scheduled for 2020 in China \cite{Lu2019}, 2021 in the US and 2023 \cite{Gagliardi2019} in Europe.
Although the details are not yet fully disclosed,
it seems that the number of nodes will not be larger than $10^5$ and will thus remain in the range of previous
machines such as the BlueGene. The major challenge will thus be to exploit the node level performance of more than 10 TFLOP/s.

The rest of this paper is organized as follows. In \replaced{S}{s}ection~\ref{sec_async} we lay
the foundations of asynchronicity and resilience, while \replaced{S}{s}ection~\ref{sec_solvers} discusses
several aspects of hardware-aware and scalable iterative linear solvers.
These building blocks will then be used in \replaced{S}{s}ections~\ref{sec_ms} and~\ref{sec_uq} to drive localized reduced basis and multilevel Monte-Carlo methods.
Finally, \replaced{S}{s}ection~\ref{sec_app} covers our surface-subsurface flow application.

\section{Asynchronicity and Fault Tolerance}
\label{sec_async}
%
%
As predicted in the first funding period, latency has indeed become a major issue,
both within a single node as well as between different
MPI ranks. The core concept underlying all latency- and communication-hiding
techniques is asynchronicity. This is also crucial to efficiently implement certain
local-failure local-recovery methods. Following the \dune philosophy, we have designed
a generic layer that abstracts the use of asynchronicity in MPI from the user.
In the following, we first describe this layer and its implementation, followed by
representative examples on how to build middleware infrastructure on it, and on its use for
s-step Krylov methods and fault tolerance beyond global checkpoint-restart techniques.

\subsection{Abstract layer for asynchronicity}\label{subsec::async_abstraction}

We first introduce a general abstraction for asynchronicity in
parallel MPI applications, which we developed for \dune. While we
integrated these abstractions with the \dune framework, most of the
code can easily be imported into other applications, and is available as a standalone library.

The C++ API for MPI was dropped from MPI-3 since it offered no real
advantage over the C bindings, beyond being a simple wrapper
layer. Most MPI users coding in C++ are still using the C bindings,
writing their own C++ interface/layer, in particular in more generic software frameworks. At the same time the C++11
standard introduced high-level concurrency concepts, in particular the
future/promise construct to enable an asynchronous program flow while
maintaining value semantics. We adopt this approach as a first
principle in our MPI layer to handle asynchronous MPI
operations and propose a highlevel C++ MPI interface, which we provide
in \dune under the generic interface of
\lstinline!Dune::Communication! and a specific implementation
\lstinline!Dune::MPICommunication!.

An additional issue of the concrete MPI library in conjunction with C++ is the
error handling concept. In C++, exceptions are the advocated approach
to handle error propagation. As exceptions change the local code path
on the, e.g., failing process in a hard fault scenario, exceptions can easily lead to a deadlock. As
we discuss later, the introduction of our asynchronous
abstraction layer enables global error handling in
an exception friendly manner.

In concurrent environments a C++ future decouples values from the actual
computation (promise). The program flow can continue while a thread is
computing the actual result and promotes this via promise to the
future. The MPI C and Fortran interfaces offer asynchonous operations,
but in contrast to thread parallel, the user does not specify the
operation within the concurrent operation. Actually,
MPI on its own does not offer any real concurrency at all, and
provides instead a handle\replaced{-}{ }based programming interface to
avoid certain cases of deadlocks: The control flow is allowed to
continue without finishing the communication, while the communication
usually only proceeds when calls into the MPI library are executed.

We developed a C++ layer on top of the asynchonous MPI operations,
which follows the design of the C++11 future. Note that the actual
\lstinline!std::future! class can not be used for this purpose.

\begin{lstlisting}
  template<typename T>
  class Future{
    void wait();
    bool ready() const;
    bool valid() const;
    T get();
  };
\end{lstlisting}

As different implementations like thread-based \lstinline!std::future!, task-based
\lstinline!TBB::future!, and our new \lstinline!MPIFuture! are available, usability greatly benefits from a
dynamically typed interface.
\deleted{interface.}
This is a reasonable approach, as \lstinline!std::future!
is using a dynamical interface already and also the MPI operations are
coarse grained, so that the additional overhead of virtual functions
calls is negligible. At the same time the user expects a future to
offer value semantics, which contradicts the usual pointer semantics
used for dynamic polymorphism. In \exadune we decided to implement
type-erasure to offer a clean and still flexible user interface.
An \lstinline!MPIFuture! is responsible for handling all state\added{s}
associated with an MPI operation.

\begin{lstlisting}
  class MPIFuture{
  private:
	mutable MPI_Request req_;
	mutable MPI_Status status_;
	impl::Buffer<R> data_;
	impl::Buffer<S> send_data_;
  public:
    ...
  };
\end{lstlisting}

The future holds a mutable \lstinline!MPI_Request! and
\lstinline!MPI_Status! to access information on the current operation
and it holds buffer objects, which manage the actual data. These
buffers offer a great additional value, as we do not access the raw
data directly, but can include data transformation and varying
ownership. For example it is now possible to directly send an
\lstinline!std::vector<double>!, where the receiver automatically resizes the
\lstinline!std::vector! according to the incoming data stream.

This abstraction layer enables different use cases, highlighted below:
\begin{enumerate}
\item \textbf
  {Parallel C++ exception handling}:
  Exceptions are the recommended way to handle faults in C++
  programs. As exceptions alter the execution path of a single node,
  they are not suitable for parallel programs. As asynchronicity
  allows for moderately diverging execution paths, we can use it to
  implement parallel error propagation using exceptions.
\item \textbf
  {Solvers and preconditioners tolerant to hard and soft faults}:
  This functionality is used for failure propagation, restoration of MPI in case of a hard fault, and asynchronous in-memory checkpointing.
\item \textbf{Asynchronous Krylov solvers}:
  Scalar products in Krylov methods require global
  communication. Asynchronicity can be used to hide the latency and
  improve strong scalability.
\item \textbf
  {Asynchronous parallel IO}: 
  The layer allows to
  transform any non-blocking MPI operation into a really asynchronous
  operation. This allows also to support asynchronous IO, to hide the
  latency of write operations and overlap with the computation of the
  next iteration or time step.
\item \textbf{Parallel Localized Reduced Basis Methods}:
  Asynchronicity will be used to mitigate the load-imbalance inherent in the
  error estimator guided adaptive online enrichement of local reduced bases.
\end{enumerate}

\subsection{Parallel C++ exception handling}\label{ssec:parallel_c++_exception_handling}

In parallel numerical algorithms, unexpected behaviour can occur quite frequently: A solver could diverge, the input of a component (e.g., the mesher) could be inappropriate for another component (e.g., the discretiser), etc.
A well-written code should detect unexpected behaviour and provide users with a possibility to react appropriately in their own programs, instead of simply terminating with some error code.
For C++, exceptions are the recommended method to handle this. With well placed exceptions and corresponding try-catch blocks, it is possible to accomplish a more robust program behaviour.
However, the current MPI specification~\cite{mpi:2015} does not define any way to propagate exceptions from one rank (process) to another. In the case of unexpected behaviour within the MPI layer itself, MPI programs simply terminate, maybe after a time-out.
This is a design decision that unfortunately implies a severe disadvantage in C++, when combined with the ideally asynchronous progress of computation and communication: An exception that is thrown locally by some rank can currently lead to a communication deadlock, or ultimately even to undesired program termination. Even though exceptions are technically an illegal use of the MPI standard (a peer no longer participates in a communication), it undesirably conflicts with the C++ concept of error handling.

Building on top of the asynchronicity layer, we have developed an approach to enable parallel C++ exceptions. We follow C++11 techniques, e.g., use future-like abstractions to handle asynchronous communication. Our currently implemented interface requires ULFM~\cite{bland2012proposal}, an MPI extension to restore communicators after rank losses, which is scheduled for inclusion into MPI-4. We also provide a fallback solution for non-ULFM MPI installations, that employs an additional communicator for propagation and can, by construction, not handle hard faults, i.e., the loss of a node resulting in the loss of rank(s) in some communicator.

To detect exceptions in the code we have extended the
\lstinline{Dune::MPIGuard}, that previously only
implemented the scope guard concept to detect
and react on local
exceptions. Our extension revokes the MPI communicator
using the ULFM functionality if a\added{n} exception is detected, so that it is
now possible to use communication inside a block with scope guard. This makes it
superfluous to call the \lstinline{finalize} and \lstinline{reactivate} methods
of the \lstinline{MPIGuard} before and after each
communication.

\begin{lstlisting}[label=lst:MPIGuard, caption={MPIGuard}]
  try{
    MPIGuard guard(comm);
    do_something();
    communicate(comm);
  }catch(...){
    comm.shrink();
    recover(comm);
  }
\end{lstlisting}
Listing \ref{lst:MPIGuard} shows an example how to use the \lstinline{MPIGuard} and recover
the communicator in a node loss scenario. In this example, an exception that is thrown only on a few ranks in
\lstinline{do_something()} will not lead to a\deleted{n} deadlock, since the \lstinline{MPIGuard} would
revoke the communicator.
Details of the implementation and further descriptions are available in a previous
publication~\cite{Engwer:2018:AHL}. We provide the ``black-channel'' fallback implementation as a
standalone version.\footnote{\url{https://gitlab.dune-project.org/exadune/blackchannel-ulfm}}
This library uses the P-interface of the MPI standard, which makes it possible
to redefine MPI functions. At the initialization of the MPI setting the
library creates an opaque communicator, called blackchannel, on which a pending
\lstinline{MPI_Irecv} request is waiting. Once a communicator is revoked, the
revoking rank sends messages to the pending blackchannel request. To avoid
deadlocks, we use \lstinline{MPI_Waitany} to wait for a request, which listens also
for the blackchannel request. All blocking communication is redirected to
non-blocking calls using the P-interface. The library is linked via
\lstinline{LD_PRELOAD} which makes it usable without recompilation and could be
removed easily once a proper ULFM implementation is available in MPI.

\begin{figure}[htb]
  \centering
    \includegraphics[width=.45\textwidth]{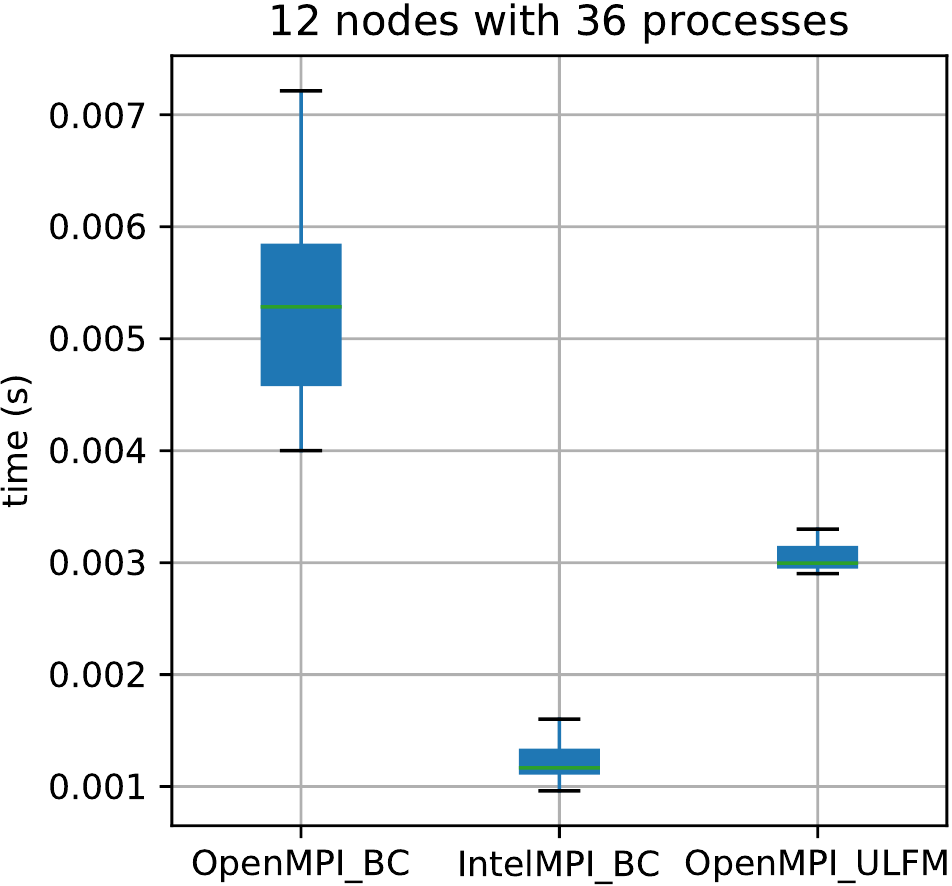}
    \hspace*{0.05\textwidth}
    \includegraphics[width=.45\textwidth]{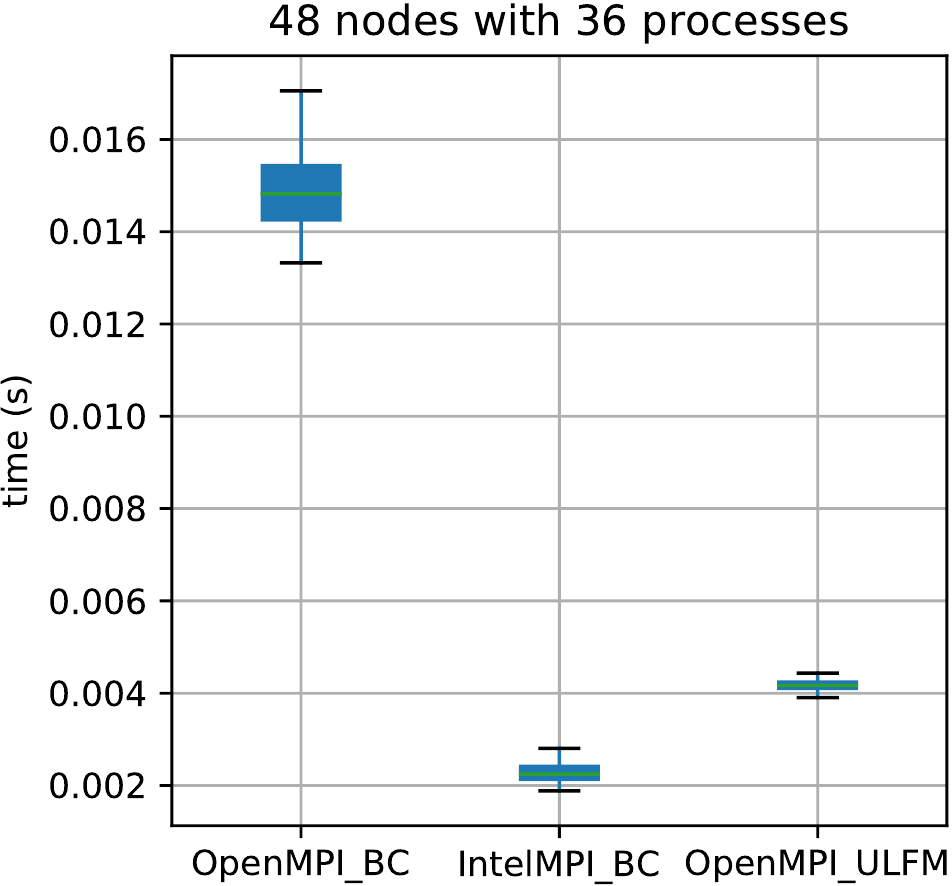}
    \caption{Benchmark of different MPI implementations: 12 nodes with 36
processes (left), 48 nodes with 36 processes (right)\added{, cf. \cite{Engwer:2018:AHL}}}
  \label{fig:blackchannel_benchmark}
\end{figure}

Figure \ref{fig:blackchannel_benchmark} shows a benchmark\deleted{s} comparing the
time which is used for duplicating a communicator, revoking it and restore a
valid state. The benchmark was performed on PALMA2, the HPC cluster of the
University of Muenster. Three implementations are compared; \textit{OpenMPI\_BC}
and \textit{IntelMPI\_BC} are using the blackchannel library based on OpenMPI
and IntelMPI, respectively. \textit{OpenMPI\_ULFM} uses the ULFM implementation
provided by \url{fault-tolerance.org}, which is based on OpenMPI. We performed
100 measurements for each implementation. The blackchannel implementation is
competitive to the ULFM implementation.

\subsection{Compressed in-memory checkpointing for linear solvers}

The previously described parallel exception propagation, rank loss detection and communicator restoration by using the ULFM extension, allow\deleted{s} us to implement a flexible in-memory checkpointing technique which has the potential to recover from hard faults on-the-fly without any user interaction. Our implementation establishes a backup and recovery strategy which in part is based on a local-failure local-recovery (LFLR)~\cite{Teranishi:2014:tlf} approach, and involves lossy compression techniques to reduce the memory footprint as well as bandwidth pressure.
\added{The contents of this subsection have not been published previously.}

\textbf{Modified solver interface.} To enable the use of exception propagation as illustrated in the previous section and to implement different backup recovery approaches we kept all necessary modifications to \istl, the linear solver library. We embed the solver initialisation and the iterative loop in a try-catch block, and provide additional entry and execution points for recovery and backup, see Listing~\ref{lst:exceptions:solver} for details. Default settings are provided on the user level, i.e., \pdelab.

\begin{lstlisting}[label=lst:exceptions:solver,numbers=left, caption={Solver modifications}]
  init_variables();
  done = false;
  while (!done) try {
    MPIGuard guard(comm);
    if (this->processRecovery(...)) /*@ \label{lst:exceptions:solver:recovery} @*/
      reinit_execution();
    } else {
      init_execution();
    }
    for (i=0 ; i<=maxit; i++ ) {
      do_iteration();
      if (converged) {
        done = true;
        break;
      }
      this->processBackup(...); /*@ \label{lst:exceptions:solver:backup} @*/
    }
  } catch(Exception & e) {
    done = false;
    comm.reconstitute(); /*@ \label{lst:exceptions:solver:reconstitute} @*/
    if (!this->processOnException(...)) /*@ \label{lst:exceptions:solver:exception} @*/
      throw;
  }
\end{lstlisting}

This implementation ensures that the iterative solving process is active until the convergence criterion is reached. An exception inside the try-block on any rank is detected by the \lstinline{MPIGuard} and propagated to all other ranks, so that all ranks will jump to the catch-block.

This catch-block can be specialised for different kind of exceptions, e.g., if a solver has diverged and a corresponding exception is thrown it could define some specific routine to define a modified restart with a possibly more robust setting and/or initial guess. The catch-block in Listing~\ref{lst:exceptions:solver} exemplarily shows a possible solution in the scenario of a communicator failure, e.g., a node loss which is detected by using the ULFM extension to MPI, encapsulated by our wrapper for MPI exceptions. Following the detection and propagation, all still valid ranks end up in the catch-block and the communicator must be re-established in some way (Listing~\ref{lst:exceptions:solver}, line~\ref{lst:exceptions:solver:reconstitute}). This can be done by shrinking the communicator or replacing lost nodes by some previously allocated spare ones. After the communicator reconstitution a user-provided stack of functions can be executed (Listing~\ref{lst:exceptions:solver}, line~\ref{lst:exceptions:solver:exception}) to react on the exception. If there is no on-exception-function or neither of them returns true the exception is re-thrown to the next higher level, e.g., from the linear solver to the application level, or in case of nested solvers, e.g. in optimisation or uncertainty quantification.

Furthermore, there are two additional entry points for user provided function stacks: In line~\ref{lst:exceptions:solver:recovery} of Listing~\ref{lst:exceptions:solver} a stack of recovery functions is executed and if it returns true, the solver expects that some modification, i.e., recovery, has been done. In this case it could be necessary that the other participating ranks have to update some data, like resetting their local right hand side to the initial values. The backup function stack in line~\ref{lst:exceptions:solver:backup} allows the user to provide functions for backup creation etc., after an iteration finished successfully.

\textbf{Recovery approaches.} First, regardless of these solver modifications, we describe the recovery concepts which are implemented into an exemplary recovery interface class providing functions that can be passed to the entry points within the modified solver. The interoperability of these components and the available backup techniques are described later. Our recovery class supports three different methods to recover from a data loss. The first approach is a global rollback to a backup, potentially involving lossy compression: Progress on non-faulty ranks may be lost but the restored data originate from the same state, i.e., iteration. This means there is no asynchronous progression in the recovered iterative process but possibly just an error introduced through the used backup technique, e.g., through lossy compression. This compression error can reduce the quality of the recovery and lead to additional iterations of the solver, but is still superior to a restart, as seen later.
For the second and third approaches, we follow the local-failure local-recovery strategy and re-initialize the data which are lost on the faulty rank by using a backup. The second, slightly simpler strategy uses these data to continue with solver iterations. The third method additionally smoothes out the probably deteriorated (because of compression) data by solving a local auxiliary problem~\cite{Goeddeke:2015:ftf,huber:2016:rfm}. This problem is set up by restricting the global operator to its purely local degrees of freedom with indices $\mathcal F \subset \mathbb N$ and a Dirichlet boundary layer. The boundary layer can be obtained by extending $\mathcal F$ to some set $\mathcal J$ using the ghost layer, or possibly the connectivity pattern of the operator $\mathbf A$. The Dirichlet values on the boundary layer are set to their corresponding values $x_N$ on the neighbouring ranks and thus additional communication is necessary:

\begin{align*}
  \mathbf{A}(\mathcal F, \mathcal F) \tilde x(\mathcal F) &= b (\mathcal F)&\text{in }\mathcal F\\
  \tilde x &= x_N& \text{on }\mathcal J \backslash \mathcal F
\end{align*}

If this problem is solved iteratively and backup data are available, the computation speed can be improved by initializing $\tilde x$ with the data from the backup.

\textbf{Backup techniques.} Our current implementation provides two different techniques for compressed backups as well as a basic class which allows \lq zero\rq-recovery (zeroeing of lost data) if the user wants to use the auxiliary solver in case of data loss without storing any additional data during the iterative procedure.

The next backup class uses a multigrid hierarchy for lossy data compression. Thus it should only be used if a multigrid operator is already in use within the solving process because otherwise the hierarchy has to be build beforehand and introduces additional overhead. Compressing the iterative vector with the multigrid hierarchy currently involves a global communication. In addition there is no adaptive control of the compression depth (i.e., hierarchy level where the backup is stored), but it has to be specified by the  user, see a previous publication for details~\cite{Goeddeke:2015:ftf}.

\begin{figure}[htb]
  \includegraphics[width=1.\textwidth,keepaspectratio]{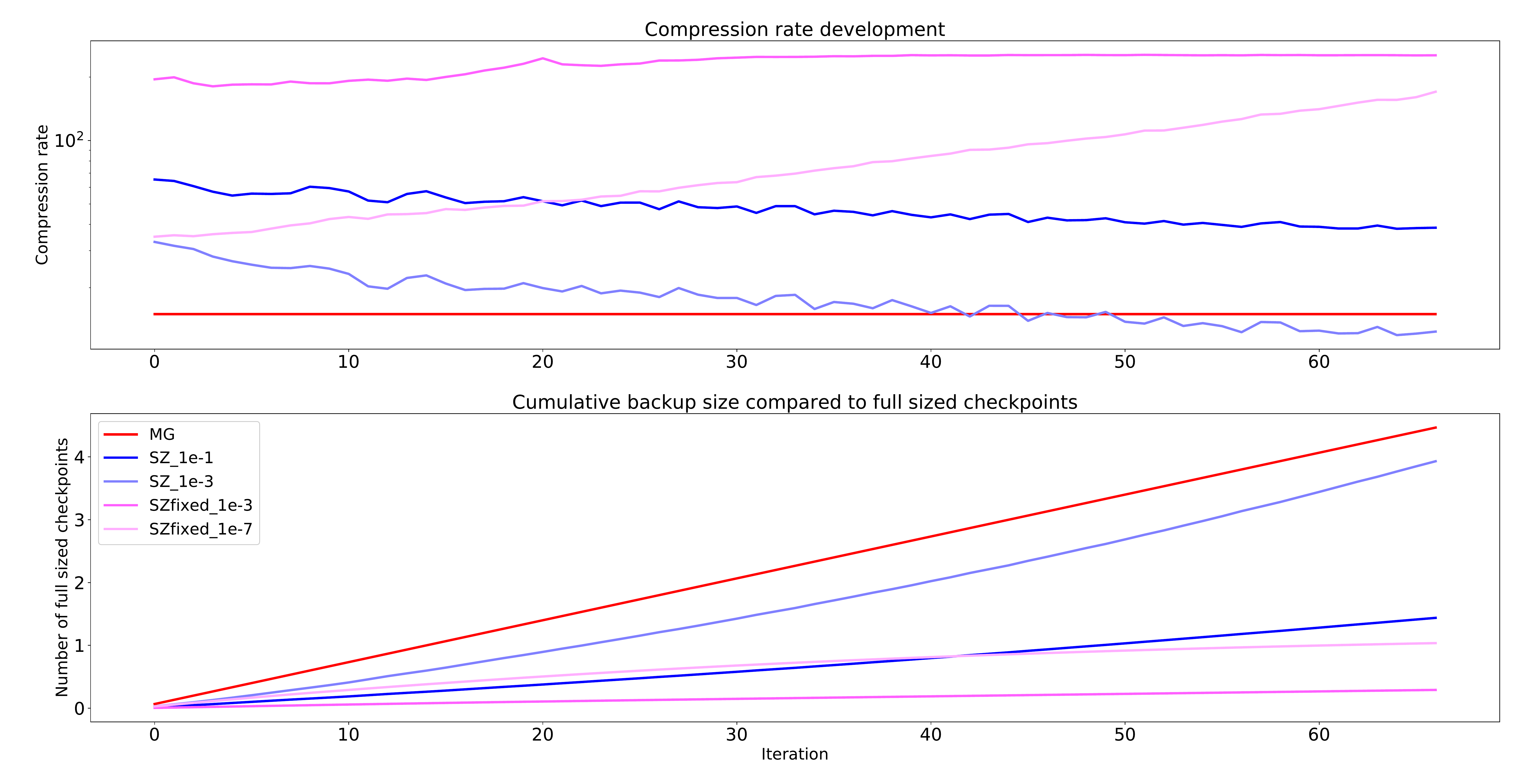}
  \caption{Compression rate in the iterative solution for an anisotropic Poisson problem on 52 cores with approximately 480\,K DOF per core}\label{fig:backup:compressionrate}
\end{figure}

\begin{figure}[htb]
  \includegraphics[width=0.9\textwidth,keepaspectratio]{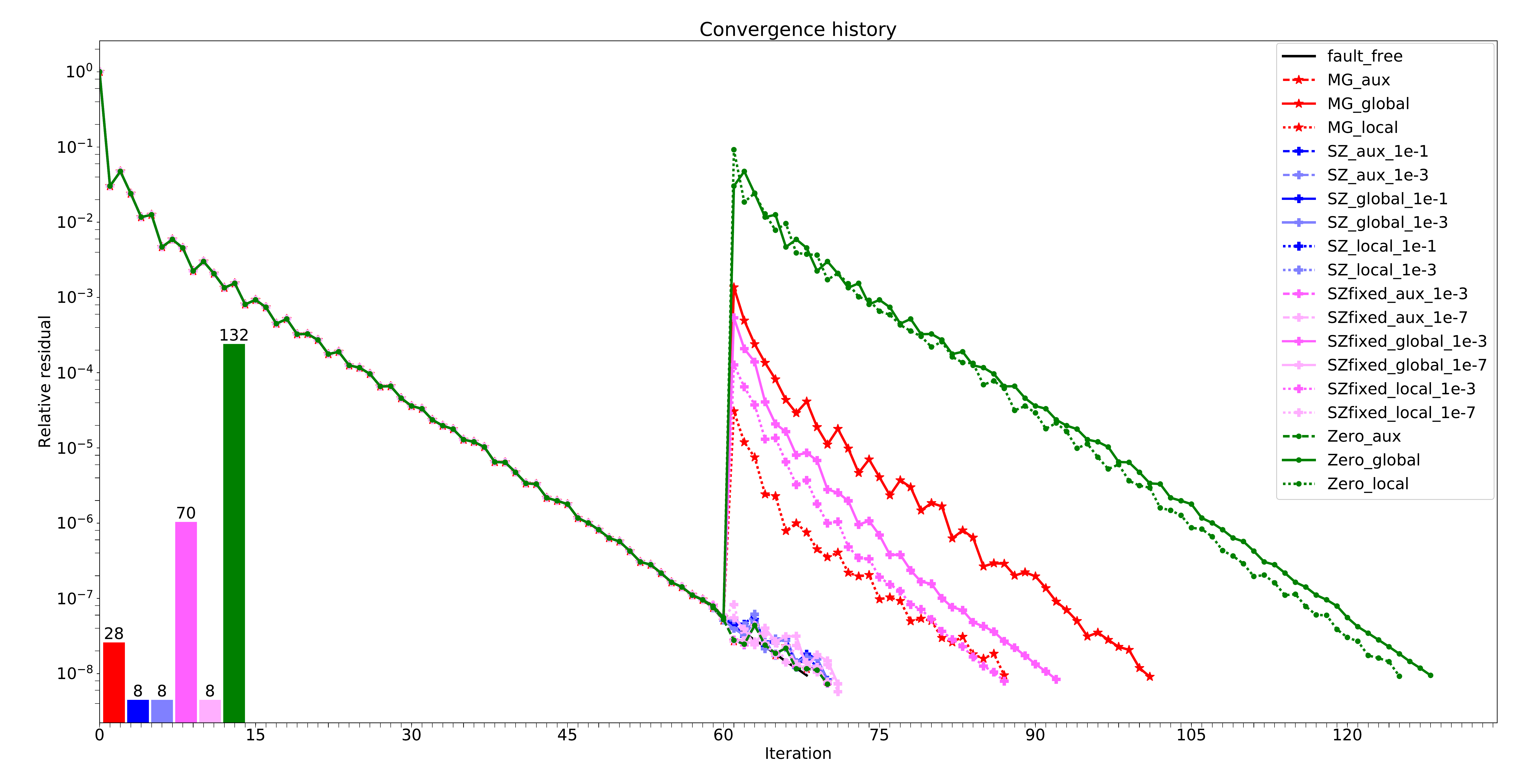}
  \caption{Convergence history in case of data loss and recovery on one rank, same setting as in Figure~\ref{fig:backup:compressionrate}. Bottom left: number of iterations to solve the auxiliary problem when using the backups as initial guess.
  \added{Note that the groups of the same colour are important, not the individual graphs.}}\label{fig:backup:convergencehistory}
\end{figure}

We also implemented a compressed backup technique based on SZ compression~\cite{Liang:2018:ecl}. SZ allows compression to a specified accuracy target and can yield better compression rates than multigrid compression. The compression itself is purely local and does not involve any additional communication. We provide an SZ backup with a fixed user-specified compression target as well as a fully adaptive one which couples the compression target to the residual norm within the iterative solver. For the first we achieve an increased rate while we approach the approximate solution, as seen in Figure~\ref{fig:backup:compressionrate} (top, pink lines), at the price of an increased overhead in case of a data loss (cf. Figure~\ref{fig:backup:convergencehistory}). The backup with adaptive compression target (blue lines) gives more constant compression rates, and a better recovery in case of faults in particular in the second half of the iterative procedure of the solver.

The increased compression rate for the fixed SZ backup is obtained because, during the iterative process, the solution gets more smooth and thus can be compressed better by the algorithm. For the adaptive method this gain is counteracted by the demand of a higher compression accuracy.

All backup techniques require to communicate a data volume smaller than the volume of four full checkpoints, see Figure~\ref{fig:backup:compressionrate} (bottom). Furthermore this bandwidth requirement is distributed over all 68 iterations (in the fault-free scenario) and could be decreased further by a lower checkpoint frequency.

The chosen backup technique is initiated before the recovery class and passed to it. Further backup techniques can be implemented by using the provided base class and overloading the virtual functions.

\textbf{Bringing the approaches together.} The recovery class provides three functions which are added to the function stacks within the modified solver interface. The backup routine is  added to the stack of backup functions of the specified iterative solver and generates backups of the current iterative solution by using the provided backup class.

\begin{lstlisting}[label=lst:recovery:interaction]
  SomeSolver solver;
  SomeBackup backup;
  Recovery recovery(backup);
  solver.addBackupFunction(&Recovery::backup, &recovery);
\end{lstlisting}

To adapt numerical as well as communication overhead for different fault scenarios and machine characteristics, the backup creation frequency can be varied. After the creation of the backup it is sent to a remote rank where it is kept in memory but never written to disk. In the following this is called \lq remote backup\rq. Currently the backup propagation happens circular by rank. It is also possible to trigger writing a backup to disk.

In the near future we will implement an on-the-fly recovery if an exception is thrown.
These will be provided to the other two function stacks and will differ depending on the availability of the ULFM extensions: If the extension is not available we can only detect and propagate exceptions but not recover a communicator in case of hard faults, i.e., node losses (cf. Section \ref{ssec:parallel_c++_exception_handling}). In this scenario the function provided to the on-exception stack will only write out the global state. Fault-free nodes will write the data of the current iterative vector, whereas for faulty nodes the corresponding remote backup is written. In the following the user will be able to provide a flag to the executable which modifies the backup object initiation to read in the stored checkpoint data. Afterwards the recovery function of our interface will overwrite the initial values of the solver with the checkpointed and possibly smoothed data like described above. 
If the ULFM extensions are available, the recovery can be realised without any user interaction: During the backup class initiation a global communication ensures that it is the first and therefore fault-free start of the parallel execution. If the process is a respawned one which replaces a lost rank, this communication is matched by a send communication created from the rank which holds the corresponding remote backup. This communication will be initiated by the on-exception function. In addition to this message the remote backup rank sends the stored compressed backup so that the respawned rank can use this backup to recover the lost data.

\replaced{So far, we have not fully implemented rebuilding the solver and preconditioner hierarchy, and the re-assembly of the local systems, in case of a node loss. This can be done with, e.g.,  message logging~\cite{Cantwell:2019:ami}, or similar}{Note that we do not consider the \lq catching up\rq\ process of a respawned rank within our work. We assume that the communication before the try-catch block of the solver is stored via, e.g., message logging~\cite{Cantwell:2019:ami}, or other} techniques which allow recomputing the individual data on the respawned rank without additional communication.

Figure~\ref{fig:backup:convergencehistory} shows the effect of various combinations of different backup and recovery techniques in case of a data loss on one rank after iteration 60. The problem is an anisotropic Poisson problem with zero Dirichlet boundary conditions which reaches the convergence criterion after 68 iterations in a fault-free scenario (black line). It is executed in parallel on 52 ranks with approximately 480\,000 degrees of freedom per rank. Thus one rank loss corresponds to a loss of around 2\% of data. For solving a conjugate gradient solver with an algebraic multigrid preconditioner is applied. In addition to the residual norm we show the number of iterations which are needed to solve the auxiliary problem when using different backups as initial guess at the bottom left.

The different backup techniques are colour-coded (multigrid: red; adaptive SZ compression: blue; fixed SZ compression: pink; no backup: green). For the SZ techniques we consider two cases, each with a different compression accuracy (fixed compression) respectively a different additional scaling coefficient (SZ). Recovery techniques are coded with different line styles: Global roll-back recovery is indicated by straight lines; simple local recovery is shown with dotted lines and if an auxiliary problem is solved to improve the quality of the recovery it is drawn with a dashed line style.
We observe that a zero recovery, multigrid compression and a fixed SZ backup with a low accuracy target are not competitive if no auxiliary problem is solved. The number of iterations needed until convergence then increases significantly. By applying an auxiliary solver the convergence can be almost fully restored (one additional global iteration) but the auxiliary solver needs a high amount of iterations (multigrid: 28; sz: 70; no backup: 132). Other backup techniques only need 8 auxiliary solver iterations. When using adaptive or very accurate fixed SZ compression the convergence behaviour can be nearly preserved even when only a local recovery or a global roll-back is applied. The adaptive compression technique has similar data overhead as the fixed SZ compression (cf. Figure~\ref{fig:backup:compressionrate}, bottom) but gives slightly better results: Both adaptive SZ compression approaches introduce only one additional iteration for all recovery approaches. For the accurate fixed SZ compression (SZfixed\_*\_1e-7) we have two additional iterations when using local or global recovery but if we apply the auxiliary solver we also have only one additional iteration until convergence.


\subsection{Communication aware Krylov solvers}
In Krylov methods
multiple scalar products per iteration must be computed. This involves global
sums in a parallel setting. As a first improvement we merged the evaluation of
the convergence criterion to the computation of a\deleted{n} scalar product. Obviously this
does not effect the computed value\added{s}, but the iteration terminates one iteration
later. However this reduces the number of global reductions per iteration from 3
to 2 and thus already saves communication overhead.

As a second step we modify the algorithm, such that only one global
communication is performed per iteration. This algorithm can also be found in the
paper of Chronopoulos and Gear \cite{chronopoulos1989s}.
Another optimization is to overlap the two scalar products with the application
of the operator and preconditioner, respectively. This algorithm was first
proposed by Gropp \cite{ghysels2014hiding}.
A fully elaborate version was then presented by Ghysels and Vanroose
\cite{ghysels2014hiding}. This version only needs one global reduction per
iteration, which is overlapped with both the application of the preconditioner
and operator. This algorithm is shown in Algorithm \ref{alg:p1rocg}.

\begin{figure}
\begin{multicols}{2}
  \centering
  \begin{algorithm}[H]
    \caption{PCG}
    \begin{algorithmic}
      \State $r_0 = b - Ax_0$
      \State $p_1 = Mr_0$
      \State
      \State $\rho_1 = \langle p_1, r_0 \rangle$
      \For{$i=1,\ldots$}
      \State $q_i = Ap_i$
      \State $\alpha_i = \langle p_i, q_i \rangle$
      \State $x_i = x_{i-1} + \frac{\rho_i}{\alpha_i}p_i$
      \State $r_i = r_{i-1} - \frac{\rho_i}{\alpha_i}q_i$
      \State $z_{i+1} = Mr_i$
      \State break if $\|r_i\| < \varepsilon$
      \State $\rho_{i+1} = \langle z_{i+1}, r_i \rangle$
      \State $p_{i+1} = \frac{\rho_{i+1}}{\rho_i} p_i + z_{i+1}$
      \EndFor
    \end{algorithmic}
    \label{alg:pcg}
  \end{algorithm}
  \columnbreak
  \begin{algorithm}[H]
  \caption{Pipelined CG}
  \begin{algorithmic}
    \State $r_0 = b - Ax_0$
    \State $p_1 = Mr_0$
    \State $q_1 = Ap_1$
    \State $\rho_1 = \langle p_1, r_0 \rangle$
    \State $\alpha_1 = \langle p_1, q_1 \rangle$
    \State $s_1 = Mq_1$
    \State $t_1 = As_1$
    \For{$i=1,\ldots$}
    \State $x_i = x_{i-1} + \frac{\rho_i}{\alpha_i}p_i$
    \State $r_i = r_{i-1} - \frac{\rho_i}{\alpha_i}q_i$
    \State break if $\|r_i\| < \varepsilon$
    \State $z_{i+1} = z_i - \frac{\rho_i}{\alpha_i}s_i$
    \State $w_{i+1} = w_i - \frac{\rho_i}{\alpha_i}t_i$
    \State $\rho_{i+1} = \langle z_{i+1}, r_i \rangle$
    \State $\tilde{\alpha}_{i+1} = \langle z_{i+1}, w_{i+1} \rangle$
    \State $\alpha_{i+1} = \frac{\alpha_i \rho_{i+1}^2}{\rho_i^2} + \tilde{\alpha}_{i+1}$
    \State $v_{i+1} = Mw_{i+1}$
    \State $u_{i+1} = Av_{i+1}$
    \State $s_{i+1} = \frac{\rho_{i+1}}{\rho_i} s_i + v_{i+1}$
    \State $t_{i+1} = \frac{\rho_{i+1}}{\rho_i} t_i + u_{i+1}$
    \State $p_{i+1} = \frac{\rho_{i+1}}{\rho_i} p_i + z_{i+1}$
    \State $q_{i+1} = \frac{\rho_{i+1}}{\rho_i} q_i + w_{i+1}$
    \EndFor
  \end{algorithmic}
  \label{alg:p1rocg}
\end{algorithm}
\end{multicols}
\end{figure}

With the new communication
interface, described above, we are able to compute multiple sums in one
reduction pattern and overlap the communication with computation. To apply these
improvements in Krylov solvers the algorithm must be adapted, such that the
communication is independent of the overlapping computation. For this adaption
we extend the \lstinline{ScalarProduct} interface by a function which can be
passed multiple pairs of vectors for which the scalar product should be
computed. The function returns a \lstinline{Future} which contains a
\lstinline{std::vector<field_type>}, once it has finished.

\begin{lstlisting}
  Future<vector<field_type>>
  dots(initializer_list<tuple<X&, X&>> pairs);
\end{lstlisting}
The function can be used in the Krylov methods like this:
\begin{lstlisting}
  scalarproduct_future = sp.dot_norm({{p,q}, {z, b}, {b,b}});
  // compute while communicate
  auto result = scalarproduct_future.get();
  field_type p_dot_q = result[0];
  field_type z_dot_b = result[1];
  field_type norm_b = std::sqrt(result[2]);
\end{lstlisting}

The runtime improvement of the algorithm strongly depends on the problem size and
on the hardware. On large systems the communication overhead makes up a large
part of the runtime. However, the maximum speedup is 3 for reducing the number
of global reductions and 2 for overlapping communication and computation,
compared to the standard version, so that a maximum speedup of 6 is
possible. The optimization also increases the memory requirements and vector
operations per iteration. An overview of runtime and memory requirements of the
methods can be found in Table \ref{tab:pipelinedcg_overview}.

\begin{table}[tb]
  \centering
  \begin{tabular}{c|c|c|c}
    & required memory & additional computational effort & global reductions\\
    \hline
    PCG & $4N$ & - & $2$\\
    Chronopoulos \& Gear & $6N$ & $1N$ & $1$ \\
    Gropp & $6N$ & $2N$ & $2$ overlapped \\
    Ghysels \& Vanroose & $10N$ & $5N$ & $1$ overlapped \\
  \end{tabular}
  \caption{Memory requirement, computational effort and global
    reductions per iteration for different versions of the preconditioned
    conjugate gradients method.}
  \label{tab:pipelinedcg_overview}
\end{table}

\begin{figure}[tb]
  \centering
  \includegraphics[width=0.65\textwidth]{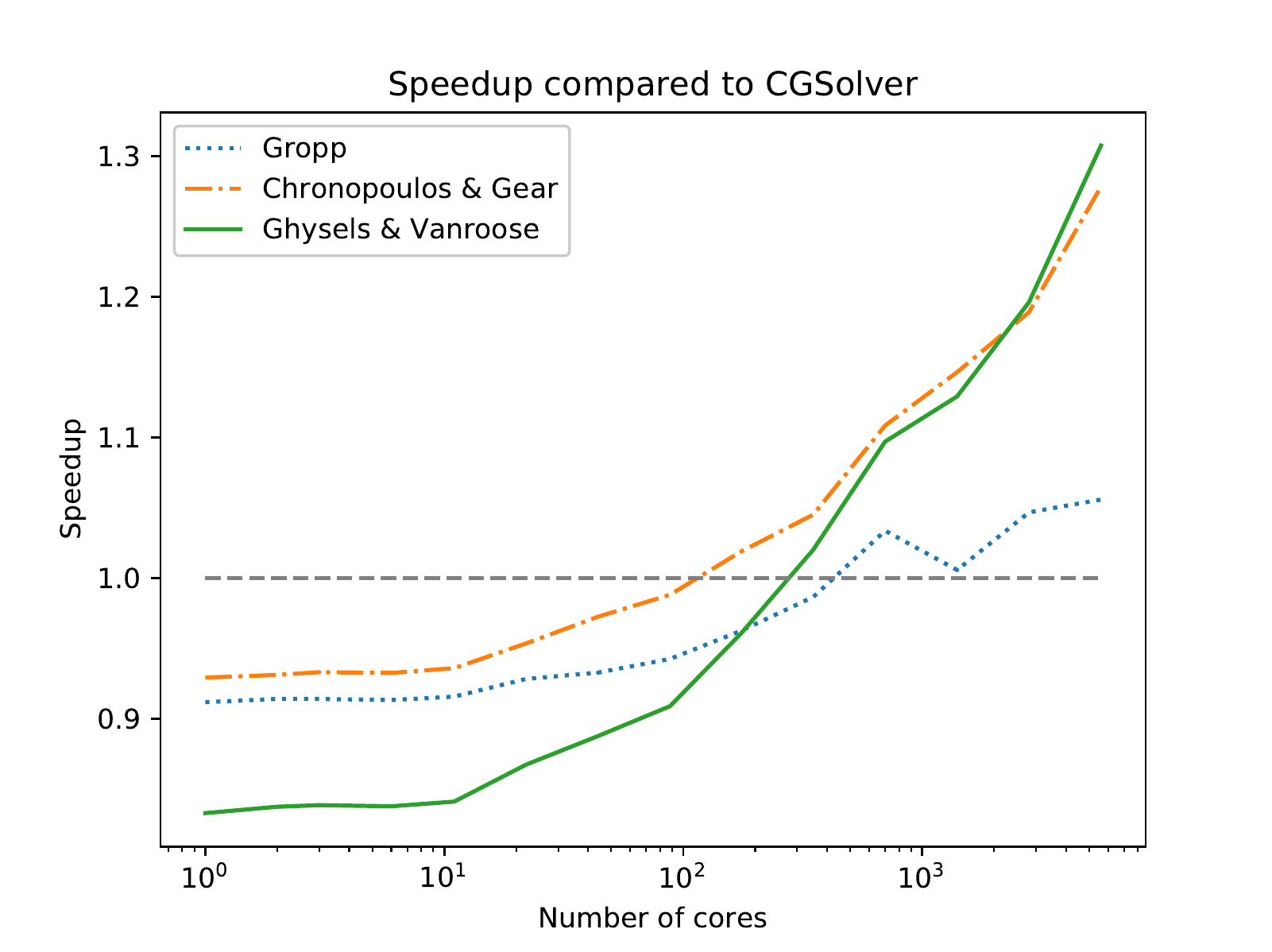}
  \caption{Strong scaling for (pipelined) Krylov subspace methods}
  \label{fig:strong_krylov_pipeline}
\end{figure}

Figure \ref{fig:strong_krylov_pipeline} shows strong scaling for different
methods. The shown speedup is per iteration and with respect to the
\lstinline{Dune::CGSolver}, which is the current CG implementation in \dune. We
use an SSOR preconditoner in an additive overlapping Schwarz setup. The problem
matrix is generated from a 5-star Finite Difference model problem. With less
cores the current implementation is faster than our optimized one. But with
higher core count our optimized versions outperforms it. The test was executed
on the helics3 cluster of the University on Heidelberg, with 5600 cores on 350
nodes. We expect that on larger systems the speedup will further increase, since
the communication is more expensive. The overlap of communication and
computation does not really come \deleted{fully} into play, since the currently used MPI
version does not support it completely.


\section{Hardware-aware, Robust and Scalable Linear Solvers}
\label{sec_solvers}
In this section we highlight improved concepts
for high-performance iterative solvers. We provide matrix-based robust solvers on GPUs
using sparse approximate inverses and optimize algorithm parameters using machine learning.
On CPUs we significantly improve the
node-level performance by using optimal matrix-free operators for Discontinous Galerkin
methods, specialized partially matrix-free preconditioners as well as vectorized linear solvers.

\subsection{Strong smoothers on the GPU: Fast Approximate Inverses with conventional and Machine Learning approaches}
In continuation of the first project phase, we enhanced the assembly of sparse approximate inverses (SPAI), a kind of preconditioner that we had shown
to be very effective within the DUNE solver before~\cite{Geveler2013327, exa:highlevel}.
Concerning the assembly of such matrices we have investigated three strategies regarding their numerical efficacy (that is their quality in approximating $A^{-1}$), the computational
complexity of the actual assembly and ultimately, the total efficiency of the amortised assembly combined with all applications during a system solution. For both strategies, this includes a decisive
performance engineering for different hardware architectures with focus on the exploitation of GPUs.

\textbf{SPAI-1.} As a starting point we have developed, implemented and tuned a fast SPAI-1 assembly routine based on MKL/LAPACK
routines (CPU) and on the cuBlas/cu\-Sparse libraries, performing up to four times faster on the GPU. This implementation is based on the
batched solution of QR decompositions that arise in Householder transformations during the SPAI minimisation process.
In many cases, we observe that the resulting preconditioner features a
high quality comparable to Gauss-Seidel methods. Most importantly, this result still
holds true when taking into account the total time-to-solution, which includes the
assembly time of the SPAI, even on a single core where the advantages of SPAI
preconditioning over forward/backward substitution during the iterative solution
process are not yet exploited. More systematic experiments with respect to these
statements as well as their extension to larger test architectures are currently being
conducted.

\begin{algorithm}[htb]
        \caption{Algorithm of the row-wise updates}
        \label{alg:rowwise}
        \begin{algorithmic}
                \For{($j=i+1,\ldots,N$)} 
                \If{$D_{jj} \neq 0$} \Comment{{\scriptsize check if the fraction is unequal
to zero}}
                \State $\alpha \gets -\frac{D_{jj}}{D_{ii}} \pmb z_i^{(i-1)}$
                \For{$n=1,\ldots,$nnz($\alpha$)}
                \If{$\alpha_n$ > $\varepsilon*max_{i,j}(A_{ij})$} \Comment{{\scriptsize here
$\alpha_n$ is the n-th entry of the vector $\alpha$}}
                \If{check($z_i^n$,$z_j^n$)} \Comment{{\scriptsize Has $z_j$ already an entry
at the columnindex of the n-th entry of $\alpha$ ?}}
                \State add($z_j^n$,$\alpha_n$)
                \State update\_minimum($z_j$) \Comment{{\scriptsize get new min. value of
j-th row}}
                \ElsIf{nnz($z_j$) < $\omega\times\frac{nnz(A)}{dim(A)}$}
\Comment{{\scriptsize maximum number of rowentries already reached?}}
                \State insert($z_j$,$\alpha_n$) \Comment{{\scriptsize insert the value
$\alpha_n$ at the fitting position}}
                \State update\_minimum($z_j$)
                \ElsIf{$\alpha_n > min(z_j)$} \Comment{{\scriptsize check if the value of
$\alpha_n$ is bigger than the minimum of $z_j$}}
                \State replace($min(z_j)$,$\alpha_n$) \Comment{{\scriptsize replace the old
minimum with the value of $\alpha_n$}}
                \State update\_minimum($z_j$)
                \EndIf
                \EndIf
                \EndFor
                \EndIf
                \EndFor
        \end{algorithmic}
\end{algorithm}

\textbf{SAINV.} This preconditioner creates an approximation of the
factorised inverse $A^{-1} = ZDR$ of a matrix $A\in\mathbb{R}^{N\times
N}$ with $D$ being a diagonal, Z an upper triangular and R a lower
triangular Matrix.

\begin{figure}[htb]
\includegraphics[width=0.95\textwidth]{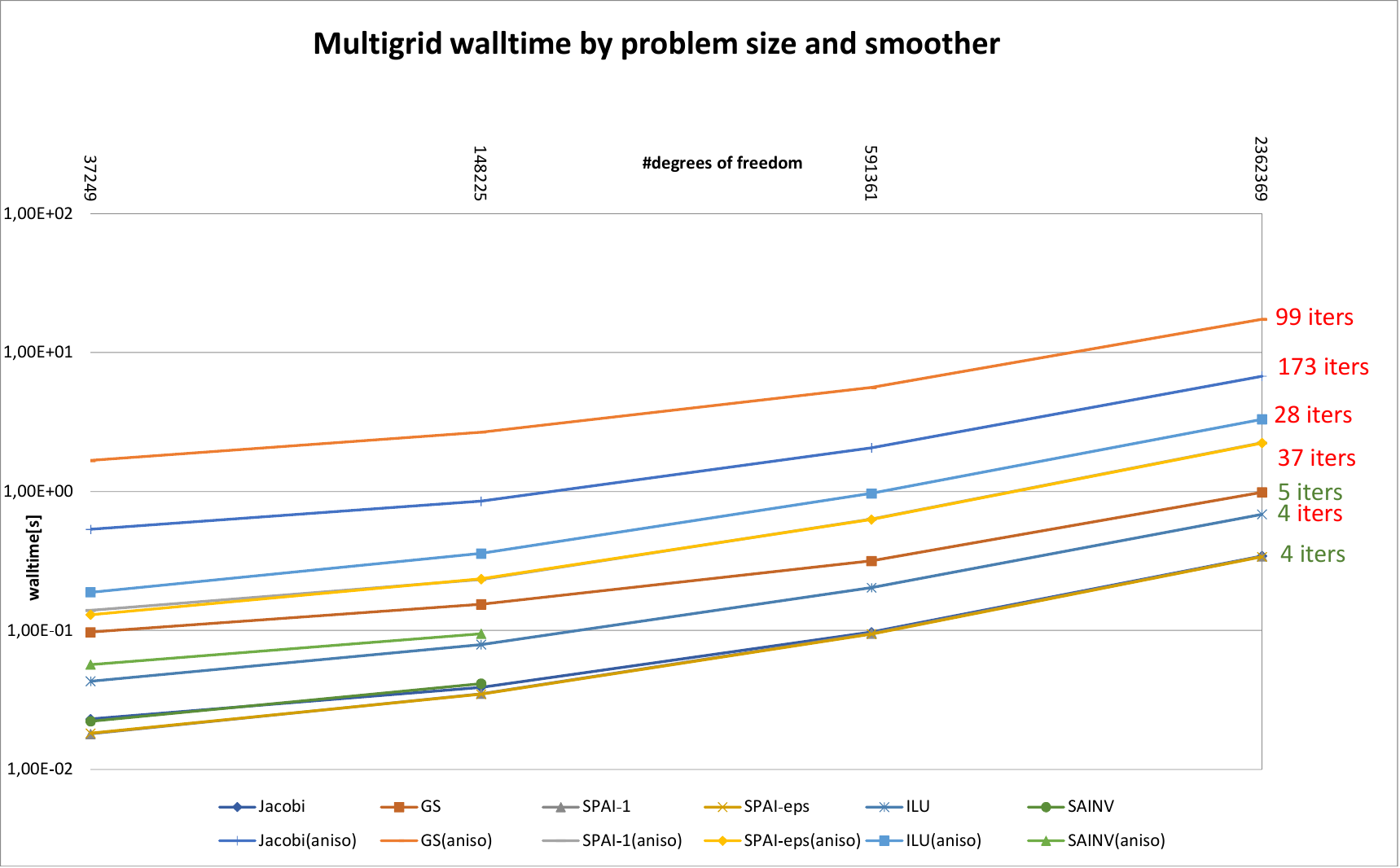}
\centering
\caption{\added{GPU smoother comparison, isotropic and anisotropic Poisson benchmarks}}\label{fig:finalresults}
\end{figure}

To describe our new GPU implementation, we write the row-wise updates in the
right-looking, outer product form of the $A$-biconjugation-process of the
SAINV factorisation as follows:
The assembly of the preconditioner is based on a loop
over the existing rows $i\in\{1,\ldots,N\}$ of $Z$ (initialised as
unit matrix $I_N$), where in every iteration the loop generally calls
three operations, namely a sparse-matrix vector multiplication, a dot
product and an update of the remaining rows $i+1,\ldots,N$ based on a
drop-parameter $\varepsilon$. \\
        In our implementation we use the ELLPACK and CSR formats, pre-allocating
a fixed amount of nonzeros of the matrix $Z$ using $\omega$ times the
average number of nonzeros per row of $A$. Having a fixed row size, no
reallocation of the arrays of the matrix format is needed and the
row-wise update can be computed in parallel. This idea is based on the
observation that while the density $\omega$ for typical drop
tolerances is not strictly limited, it generally falls into the interval
$]0,3[$.
        As the SpMV and the dot kernels are well established, we take
a closer look at the row-wise update, which is described more detailed
in algorithm \ref{alg:rowwise}. We first compute the
values to be added and store them in a variable $\alpha$. Then we
iterate over all nonzero entries of $\alpha$ (which of course has the
same sparsity pattern as $z_i$) and check if the computed value
exceeds a certain drop-tolerance. If this condition is met, we have
three conditions for an insertion into the matrix $Z$:
        \begin{enumerate}
                \item check if there is already an existing nonzero value in the $j$-th
row at the column index of the value $\alpha_n$ and search for the new
minimal entry of this row
                \item else check if there is still place in the $j$-th row, so we can
simply insert the value $\alpha_n$ into that row and search for the new
minimal entry of this row.
                \item else check if the value $\alpha_n$ is greater than the current
minimum. If this condition is satisfied, then switch the old minimal value
with $\alpha_n$ and search for the new minimal entry of this row.
        \end{enumerate}
        If none of these conditions is met, we drop the computed value without
updating the current column and repeat these steps for the next values
unequal to zero of the current row. This cap of values for each row also
has the following disadvantages:
By having a too small maximum of nonzeros per row, a qualitative
$A$-orthogonalization cannot be performed. To avoid this case we only
take values of $\omega$ greater than one, which seem\added{s} to be sufficient.
Also, if a row has already reached the maximum number of nonzeros, additional but relatively small values may be dropped. This can become an issue if the sum of these small numbers leads to a relevant entry in a later iteration.
\replaced{For a comparison, Figure~\ref{fig:finalresults} depicts the time-to-solution for V-cycle multigrid using different strong smoothers on a P100 GPU. All smoothers are constructed using 8 Richardson iterations with (reasonably damped if necessary) preconditioners such as Jacobi, Gauss-Seidel, ILU-0, SPAI-1, SPAI-$\epsilon$ and SAINV. We set up the benchmark case from a 2D Poisson problem in the isotropic case and with two-sided anisotropies in the grid to harden the problem even for well-ordered ILU approaches. The SPAI approaches are the best choice for the smoother on the GPU.
}{The benchmarks below show that the assembly phase
of SAINV can efficiently be performed on the GPU with a competitive
performance to the SPAI-$\epsilon$ assembly.}

\textbf{Machine Learning.} Finally we started investigating how to construct approximate inverses using methods from Machine Learning~\cite{RuelmannGevelerTurek2018a}. The basic idea here is to treat $A^{-1}$ as a discrete function in the course of a function regression process. The neural network therefore learns
how to deduct (extrapolate) an approximation of the inverse.
Once trained with many data pairs of matrices and their inverse (a sparse representation of it) a neural network
like a multilayer perceptron can be able to approximate inverses rapidly.
As a starting point we have employed the finite element method for the Poisson equation on
different domains with linear basis functions and have used it to generate expedient systems
of equations to solve. Problems of this kind are usually based on sparse M-matrices
with characteristics that can be used to reduce the calculation time and effort of the
neural network training and evaluation.
Our results show that given the pre-defined quality of the preconditioner (equivalent to the $\epsilon$ in a SPAI-$\epsilon$ method), we can by far numerically outperform even Gauss-Seidel.
Using Tensorflow~\cite{tensorflow} and numpy~\cite{ascher.dubois.hinsen.hugunin.oliphant-1999-np}, the learning algorithm can even be performed on the GPU.
Here we have used a three-layered fully-connected perceptron with fifty neurons in
each layer plus input and output layers, and employed the resulting preconditioners in a Richardson method to solve the
mentioned problem on a three times refined L-domain with a fixed number of degrees of freedom. The
numerical effort of each evaluation of the neural network is basically the effort of
a matrix-vector-multiplication for each layer in which the matrix size depends on
the number of neurons per layer (M) and the non zero entries (N) of the input
matrix, like $O(NM)$ for the first layer. The inner layers effort, without
input and output layer, just depends on the number of neurons.
The crucial task now is to balance the quality of the resulting approximation and the effort to evaluate the network.
We use fully connected feed-forward multilayer perceptrons as a
starting point. Fully connected means that every neuron in the network
is connected to each neuron of the next layer. Moreover there are no
backward connections between the different layers (feed-forward). The
evaluation of such neural networks is a sequence of chained matrix-vector products.

\begin{figure}[htb]
\includegraphics[height=3cm, width=8.5cm]{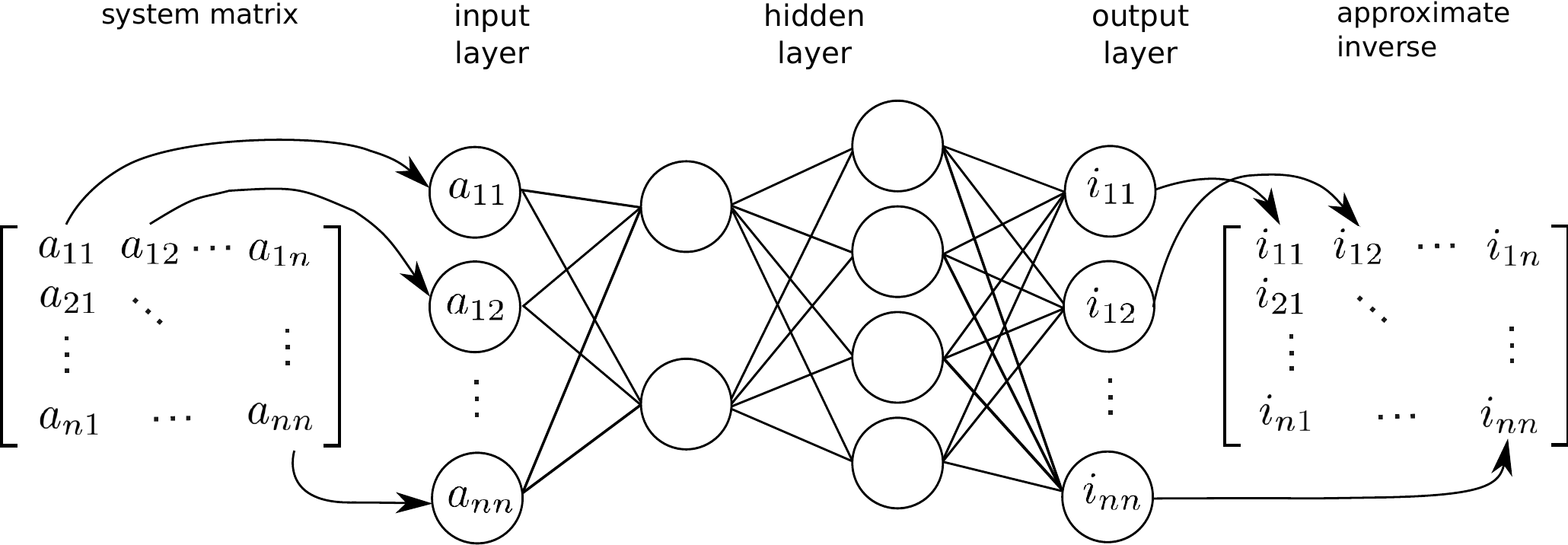}
\centering
\caption{Model of a neural network for matrix inversion\added{, cf.~\cite{RuelmannGevelerTurek2018a}}}\label{fig:neural_network}
\end{figure}

The entries of the system matrix are represented vector-wise in the input layer
(cf. Figure~\ref{fig:neural_network}). In the same way,
our output layer contains the entries of the approximate
inverse. Between these layers we can add a number of hidden layers
consisting of hidden neurons. How many hidden neurons we
need to create strong approximate inverses is a key design decision
and we discuss this below.
In general our supervised training algorithm is a backward
propagation with random initialisation. Alongside a linear propagation
function $i_{\text{total}} = \textbf{W} \cdot o_{\text{total}} + b\nonumber$
with the total (layer) net input $i_{\text{total}}$, the weight matrix $\textbf{W}$, the vector for the bias weights $b$ and the total output of the previous layer $o_{\text{total}}$, we use the rectified linear unit (ReLu) function as activation function $\alpha (x)$ and thus we can calculate the output $y$ of each neuron as $y := \alpha ( \sum_\text{j} o_\text{j} \cdot w_{\text{ij}} )$.
Here $o_\text{j}$ is the output of the preceding sending units and $w_{\text{ij}}$ are the corresponding weights between the neurons.

\begin{figure}[htb]
\includegraphics[width=0.9\textwidth]{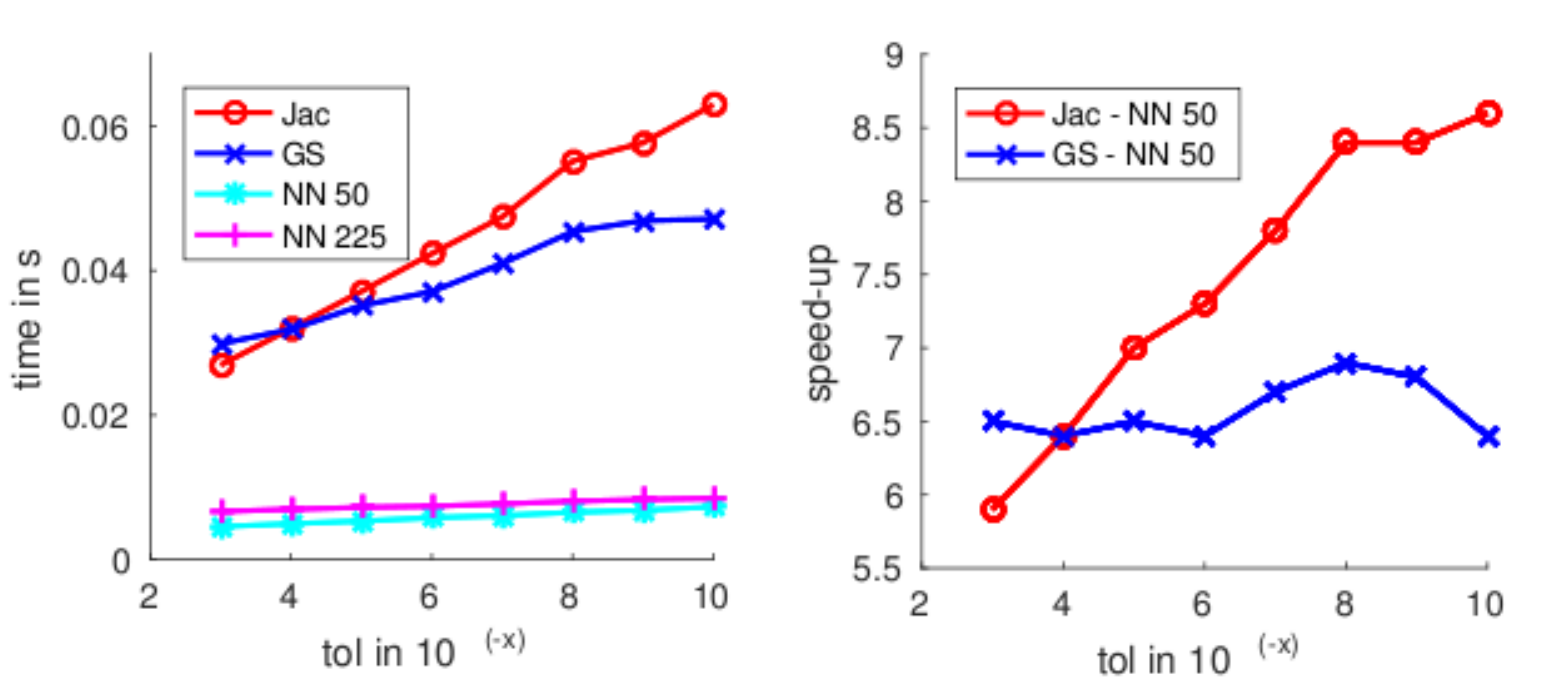}
\centering
\caption{Results for the defect correction with the neural network, cf.~\cite{RuelmannGevelerTurek2018a}}\label{fig:neural_network_results}
\end{figure}

For the optimization we use the L2 error function and update the weights with $w_{\text{ij}}^{(t+1)} = w_{\text{ij}}^{(t)} + \gamma \cdot o_i \cdot \delta_\text{j}\nonumber$, with the output $o_i$ of the sending unit and learning rate $\gamma$. $\delta_\text{j}$ symbolises the gradient decent method:
  \[
     \delta_\text{j} = \left\{\begin{array}{ll}  f'(i_\text{j}) \cdot (\hat o_\text{j} - o_\text{j})  & \text{if neuron j is an output neuron} \\
         f'(i_\text{j}) \cdot \sum_{k \in S} (\delta_k \cdot w_{kj})  & \text{if neuron j is a hidden neuron.} \end{array}\right.
  \]
For details concerning the test/training algorithm we refer to a previous publication~\cite{RuelmannGevelerTurek2018a}.
For the defect correction prototype, we find a significant speedup for a moderately anisotropic Poisson problem, see
Figure~\ref{fig:neural_network_results}.

\deleted{For a final comparison, Figure~\ref{fig:finalresults} depicts the time-to-solution for V-cycle multigrid using
  different strong smoothers on a P100 GPU. All smoothers are constructed using 8 Richardson iterations with (reasonably
  damped if necessary) preconditioners such as Jacobi, Gauss-Seidel, ILU-0, SPAI-1, SPAI-$\epsilon$ and SAINV. As above,
  we set up the benchmark case from a 2D Poisson problem in the isotropic case and with two-sided anisotropies in the
  grid to harden the problem even for well-ordered ILU approaches. The SPAI approaches are the best choice for the
  smoother on the GPU.}

\subsection{Autotuning with artificial neural networks}
\begin{figure}[htb]
  \centering
\includegraphics[width=\textwidth]{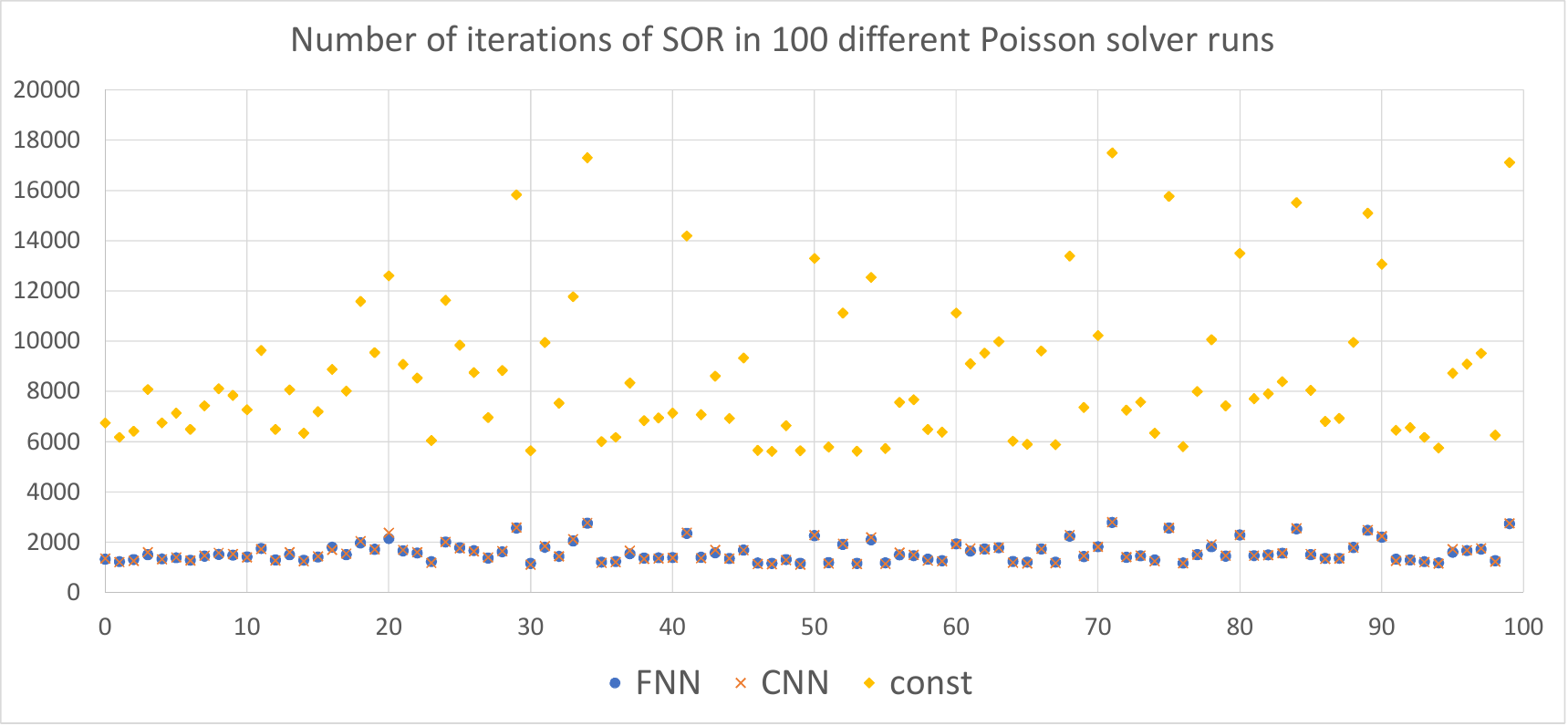}
\caption{Result \added{for 100 samples} of the FNN-based autotuning system for the $\omega$ parameter in SOR}\label{sornn}
\end{figure}

Inspired by our usage of Approximate Inverses generated by artificial neural networks (ANNs), we exploit (Feed Forward-) neural networks (FNN) for the automatic tuning of solver parameters. We were able to show that it is possible to use such an approach to provide much better a-priori choices for the parametrisation of iterative linear solvers. In detailed studies for 2D Poisson problems we conducted benchmarks for many test matrices and autotuning systems using FNNs as well as convolutionary neural networks (CNNs) to predict the $\omega$ parameter in a SOR solver. In Figure~\ref{sornn} we depict \replaced{100 randomly choosen samples}{some results} of this study. It can be seen that even for good a-priori choices of $\omega$ the NN-driven system can compete whilst \lq bad\rq\ choices (labeled constant) might lead to a stalling solver.

\subsection{Further development of sum-factorized matrix-free DG methods}
\label{subsec:matfreedg}

While we were able to achieve good node-level performance with our matrix-free DG methods in the first funding period,
our initial implementations still did not utilize more than about 10\% of the theoretical peak FLOP throughput. In the
second funding period, we systematically improved on those results by focusing on several aspects of our implementation:

\textbf{Introduction of block-based DOF processing.} Our implementation is based on \pdelab{}, a very flexible
discretization framework for both continuous and discontinuous discretizations of PDEs. In order to support a wide range
of discretizations, PDELab has a powerful system for mapping DOFs to vector and matrix entries. Due to this flexibility,
the mapping process is rather expensive. On the other hand, Discontinuous Galerkin values will always be blocked in a
cell-wise manner. This can be exploited by only ever mapping the first degree of freedom associated with each cell and
then assuming that all subsequent values for this cell are directly adjacent to the first entry. We have added a
special \lq DG codepath\rq\ to \pdelab{} which implements this optimization.

\textbf{Avoiding unnecessary memory transfers.} As all of the values for each cell are stored in consecutive locations
in memory, we can further optimize the framework behavior by skipping the customary gather / scatter steps before and
after the assembly of each cell and facet integral. This is implemented by replacing the data buffer normally passed to
the integration kernels with a dummy buffer that stores a pointer to the first entry in the global vector /
matrix and directly operates on the global values. This is completely transparent to the integration kernels, as they
only ever access the global data through a well-defined interface on these buffer objects. Together with the previous
optimization, these two changes have allowed us to reduce the overhead of the framework infrastructure on assembly times
from more than 100\% to \replaced{less than 5\%}{just a few percent}.

\textbf{Explicit vectorization.} The DG implementation used in the first phase of the project was written as scalar code
and relied on the compiler's auto vectorization support to utilize the SIMD instruction set of the processor, which we
tried to facilitate by providing compile time loop bounds and aligned data structures. In the second phase, we have
switched to explicit vectorization with a focus on AVX2, which is a common foundation instruction set across all current
x86-based HPC processors. We exploit the possibilities of our C++ code base and use a well-abstracted library which
wraps the underlying compiler intrinsic calls \cite{Fog:Vectorclass}. In a separate project \cite{2018arXiv181208075K},
we are extending this functionality to other SIMD instruction sets like AVX512.

\textbf{Loop reordering and fusion.} While vectorization is required to fully utilize modern CPU architectures, it is
not sufficient. We also have to feed the execution units with a substantial number of mutually independent chains of
computation ($\approx$ 40--50 on current CPUs). This amount of parallelism can only be extracted from typical DG
integration kernels by fusing and reordering computational loops. In contrast to other implementations of matrix-free DG
assembly \cite{Fehn:2018,Luporini2017}, we do not group computations across multiple cells or facets, but instead across
quadrature points and multiple input/output variables. In 3D, this works very well for scalar PDEs that contain both the
solution itself and its gradient, which adds up to four quantities that exactly fit into an AVX2 register.

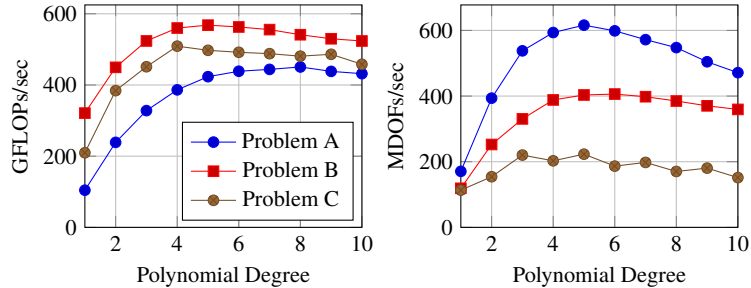
\begin{figure*}[tbh]
  \centering
  \begin{tikzpicture}
    \begin{axis}[
      name=plot1,
      width=.45\textwidth,
      xlabel={Polynomial Degree},
      ylabel={GFLOPs/sec},
      ymin=0,
      xmin=1,
      xmax=10,
      grid=major,
      legend entries={{Problem A},{Problem B},{Problem C}},
      legend pos=south east
      ]
      \addplot table[x=degree,y=flopspersec-a] {dg/performance.txt};
      \addplot table[x=degree,y=flopspersec-b] {dg/performance.txt};
      \addplot table[x=degree,y=flopspersec-c] {dg/performance.txt};
    \end{axis}
    \begin{axis}[
      at=(plot1.right of south east), anchor=left of south west,
      width=.45\textwidth,
      xlabel={Polynomial Degree},
      ylabel={MDOFs/sec},
      ymin=0,
      xmin=1,
      xmax=10,
      grid=major,
      ]
      \addplot table[x=degree,y expr=\thisrow{dofspersec-a}/1e6] {dg/performance.txt};
      \addplot table[x=degree,y expr=\thisrow{dofspersec-b}/1e6] {dg/performance.txt};
      \addplot table[x=degree,y expr=\thisrow{dofspersec-c}/1e6] {dg/performance.txt};
    \end{axis}
  \end{tikzpicture}
  \caption{Floating point performance in GFLOPs/sec and throughput in MDOFs/sec for full operator application, 2x Intel Xeon E5-2698v3 2.3
    GHz for all model problems}
  \label{fig:problem-comparison}
\end{figure*}

\textbf{Results.} Table~\ref{fig:throughput-table} compares the throughput and the hardware efficiency of our matrix-free code for two
diffusion-reaction problems A (axis-parallel grid, constant coefficients per cell) and B (affine geometries, variable
coefficients per cell) with a matrix-based implementation. Figure~\ref{fig:problem-comparison} compares throughput and
floating point performance of our implementation for these problems as well as an additional problem C
with multi-linear geometries, demonstrating that we are able to achieve more than 50\% of theoretical peak FLOP rate on
this machine as well as a good computational processing rate as measured in DOFs/sec.

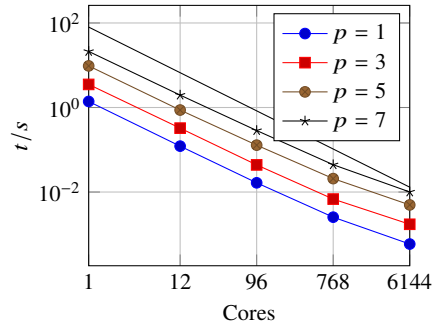
\begin{figure}[tbh]
  \centering
  \begin{tikzpicture}
    \begin{loglogaxis}[
      width=.5\textwidth,
      grid=major,
      xmin=1,
      xmax=6144,
      ylabel shift=-.4em,
      xtick={1,12,96,768,6144},
      xticklabels={1,12,96,768,6144},
      legend entries={{$p=1$},{$p=3$},{$p=5$},{$p=7$}},
      legend pos=north east,
      xlabel={Cores},
      ylabel={$t/s$},
      ]
      \addplot table[x=degree,y=Q1] {dg/strong-scalability.txt};
      \addplot table[x=degree,y=Q3] {dg/strong-scalability.txt};
      \addplot table[x=degree,y=Q5] {dg/strong-scalability.txt};
      \addplot table[x=degree,y=Q7] {dg/strong-scalability.txt};
      \addplot [domain=1:6144] {80/x};
    \end{loglogaxis}
  \end{tikzpicture}
  \caption{Runtimes for strong scalability on IWR compute cluster (416 nodes with 2 x E5-2630 v3 each, 64 GiB / node, QDR Infiniband)}
  \label{fig:strong-scaling-fig}
\end{figure}

While our work in this project was mostly focused on scalar diffusion-advection-reaction problems, we have also applied
the techniques shown here to projection-based Navier-Stokes solvers \cite{PIATKOWSKI2018220}. One important lesson
learned was the unsustainable amount of work required to extend our approach to different problems and / or hardware
architectures. This has lead us to develop a Python-based code generator in a new project \cite{2018arXiv181208075K},
which provides powerful abstractions for the building blocks listed above. This toolbox can be extended and combined in
new ways \deleted{(driven by, e.g., autotuning)} to achieve performance comparable to hand-optimized code.
\added{Especially for more complex problems involving systems of equations, there are a large number of possible ways to
  group variables and their derivatives into sum factorization kernels due to our approach of vectorizing over multiple
  quantities within a single cell. The resulting search space is too large for manual exploration, which the above
  project solved by the addition of benchmark-driven automatic comparison of those variants.} Finally, initial results
show good scalability of our code as shown by the strong scaling results in Figure \ref{fig:strong-scaling-fig}. Our
implementation shows good scalability until we reach a local problem size of just 18 cells, where we still need to
improve the asynchronicity of ghost data communication and assembly.

\begin{table*}[htb]
  \centering
  \footnotesize
\begin{tabularx}{0.78\textwidth}{cYGYGYFYG}
\toprule
& \multicolumn{2}{c}{Matrix-free A}& \multicolumn{2}{c}{Matrix-free B} & \multicolumn{2}{c}{Matrix-based} & \multicolumn{2}{c}{Matrix Assembly} \\ \cmidrule(lr){2-3} \cmidrule(lr){4-5} \cmidrule(lr){6-7} \cmidrule(lr){8-9}
$p$ & \multicolumn{1}{c}{\si[per-mode=fraction]{\dof\per\second}} & \multicolumn{1}{c}{\si[per-mode=fraction]{\giga\flop\per\second}} & \multicolumn{1}{c}{\si[per-mode=fraction]{\dof\per\second}} & \multicolumn{1}{c}{\si[per-mode=fraction]{\giga\flop\per\second}} & \multicolumn{1}{c}{\si[per-mode=fraction]{\dof\per\second}} & \multicolumn{1}{c}{\si[per-mode=fraction]{\giga\flop\per\second}} & \multicolumn{1}{c}{\si[per-mode=fraction]{\dof\per\second}} & \multicolumn{1}{c}{\si[per-mode=fraction]{\giga\flop\per\second}} \\
\specialrule{\lightrulewidth}{\belowrulesep}{\belowrulesep}
1 & 170158953.810 & 104.158182879 & 118979587.729 & 321.037185434 & 206451612.9 & 24.3316669935 & 16039016.302 & 344.537537152 \\
2 & 392974786.299 & 238.415477889 & 252483269.110 & 449.598013872 & 64200044.85 & 27.2839491589 & 8709205.2273 & 371.308869533 \\
3 & 537750267.971 & 327.958675764 & 330453991.322 & 523.700278739 & 26851310.34 & 29.8372413793 & 4662890.30518 & 368.195374202 \\
4 & 595014243.273 & 387.318881958 & 388214296.397 & 559.995679261 & 9544347.82 & 23.4740484429 & 2574016.32326 & 300.728647745 \\
5 & 617156313.613 & 424.131804157 & 403296404.441 & 568.113051526 & 4584939.03 & 23.3417548171 & 1925972.94389 & 307.258859483 \\
6 & 598962766.081 & 438.799707698 & 406075144.913 & 562.979680371 & 2310736.84 & 23.4839329435 & 1210878.64306 & 230.908894365 \\
7 & 570054393.153 & 442.435533605 & 398006289.990 & 555.586463942 & 1462449.09 & 30.3253444017 & 664872.60321 & 142.838982665 \\
8 & 540645570.859 & 444.910841687 & 384895618.842 & 541.184528147 & 614493.65 & 24.0574267317 & 873503.72048 & 197.904571983 \\
9 & 504835956.761 & 438.602901953 & 370258969.165 & 529.864853667 & 297674.41 & 26.1953488372 & 700825.043546 & 132.758132756 \\
10 & 471259916.574 & 431.763201629 & 359375741.589 & 523.586101106 & \multicolumn{1}{c}{---} & \multicolumn{1}{c}{---} & \multicolumn{1}{c}{---} & \multicolumn{1}{c}{---} \\
\bottomrule
\end{tabularx}
\caption{Full operator application, 2x Intel Xeon E5-2698v3 2.3 GHz (32 cores), for two
  problems of different complexity, and for comparison matrix assembly for the simpler problem and
  matrix-based operator application. Note that the matrix-based computations use significantly
  smaller problem sizes due to memory constraints. \added{$p$ denotes the polynomial degree of the DG space.}}
\label{fig:throughput-table}
\end{table*}

\subsection{Hybrid solvers for Discontinuous Galerkin schemes}

In Section \ref{subsec:matfreedg} we concentrated on the performance of matrix-free
\textit{operator application}. This is sufficient for instationary problems with explicit time integration,
but in case of stationary problems or implicit time integration, (linear) algebraic
systems need to be solved. This requires operator application and robust, scalable preconditioners.

For this we extended hybrid AMG-DG preconditioners \cite{amg4dg} in a joint work with Eike Müller
from Bath University, UK, \cite{BASTIAN2019417}. In a solver for matrices arising from
higher order DG discretizations the basic idea is to perform all computations on the DG system in
a matrix-free fashion and to explicitly assemble only a matrix in a low-order subspace which
is significantly smaller. In the sense of subspace correction methods \cite{Xu92} we employ
a splitting $$V_{DG}^p = \sum_{T\in\mathcal{T}_h} V_T^p + V_c$$
where $V_T^p$ is the finite element space of polynomial degree $p$ on element $T$ and
the coarse space $V_c$ is either the lowest-order conforming finite element space $V_h^1$ on the mesh $\mathcal{T}_h$,
or the space of piecewise constants $V_h^0$. Note that the symmetric weighted interior
penalty DG method from \cite{Ern01042009} reduces
to the cell-centered finite volume method with two-point flux approximation on $V_h^0$. Note also, that the system on
$V_c$ can be assembled without assembling the large DG system.

\begin{figure}[htb]
\begin{minipage}{0.45\linewidth}
\includegraphics[width=\linewidth]{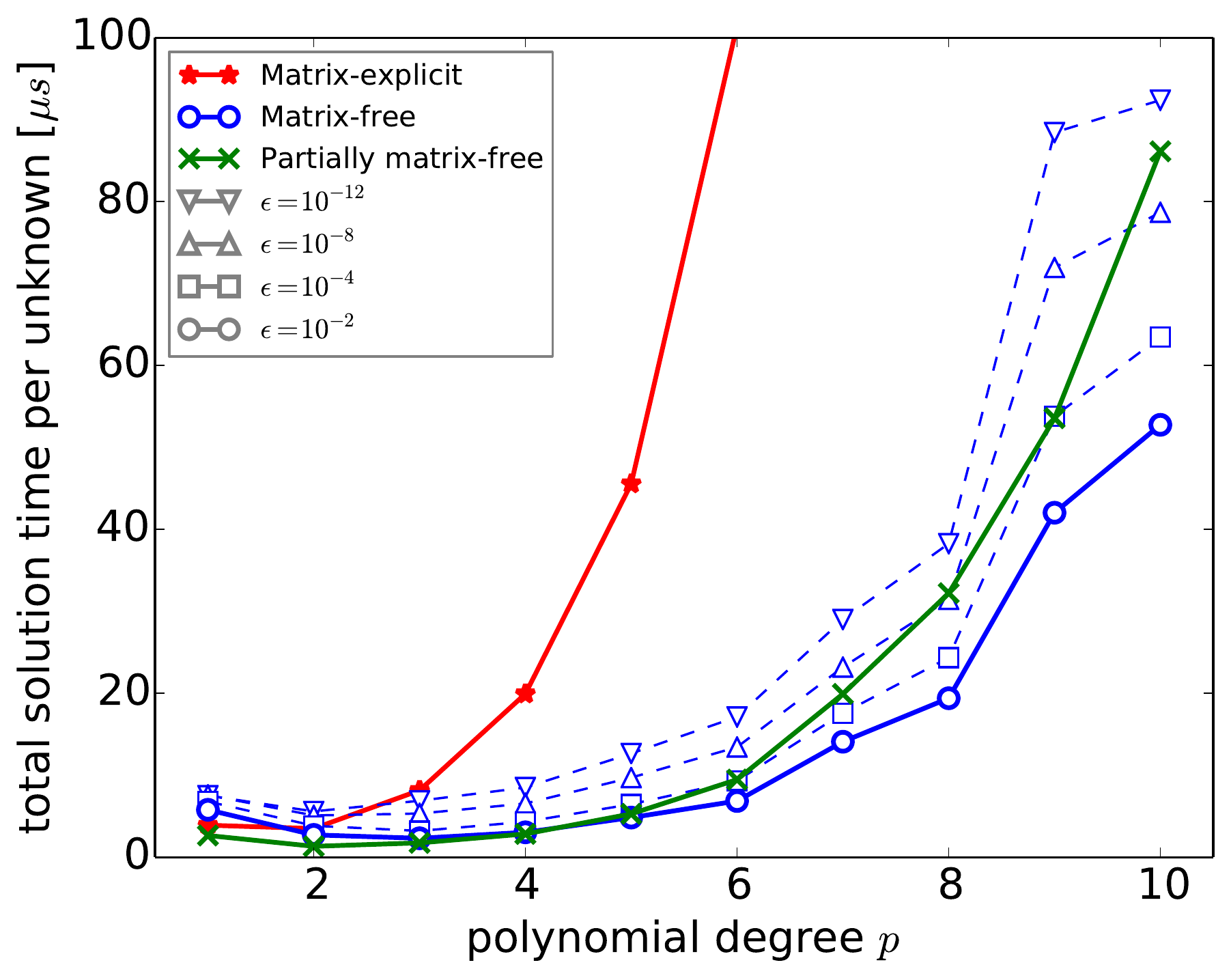}
\end{minipage}
\hfill
\begin{minipage}{0.45\linewidth}
\includegraphics[width=\linewidth]{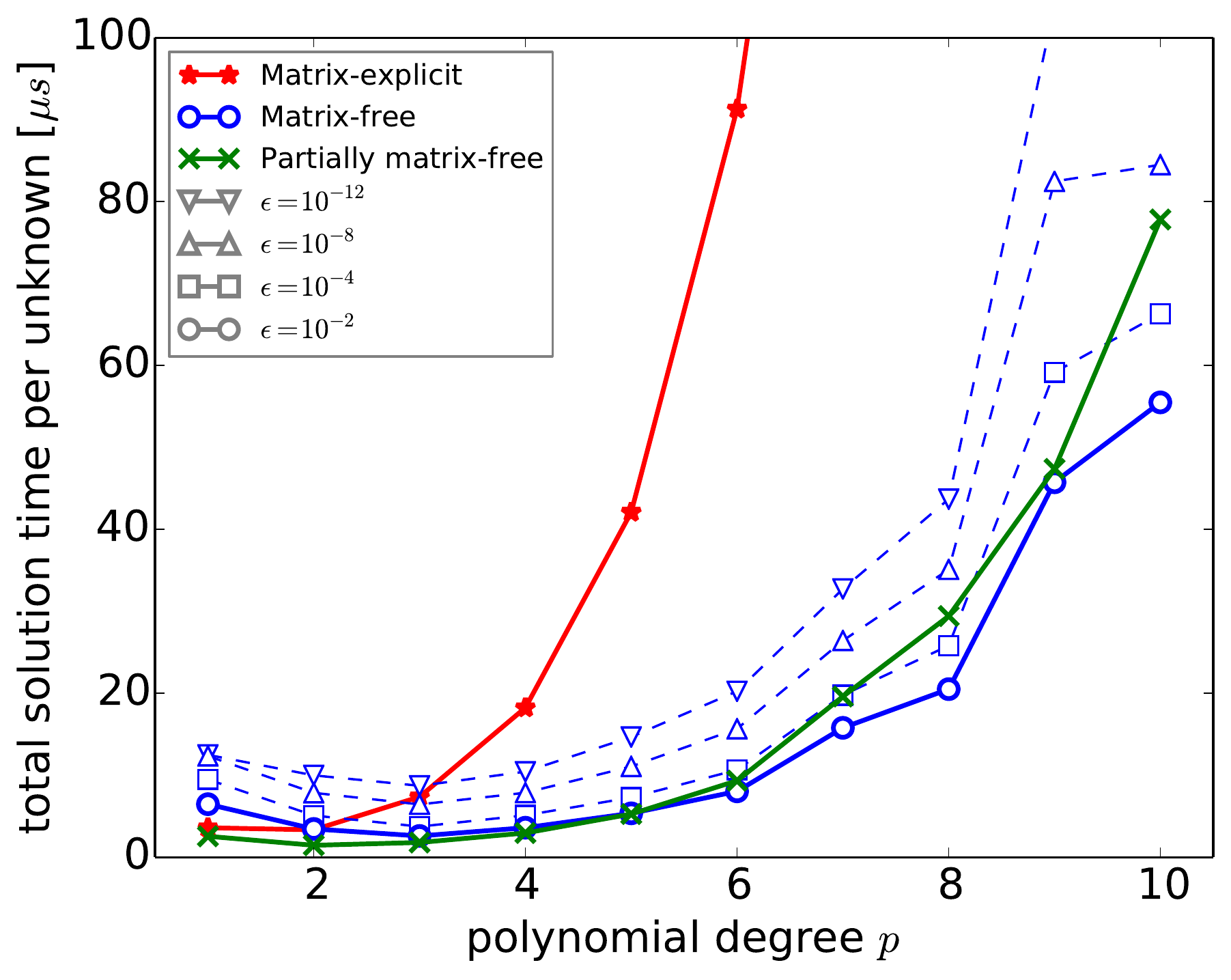}
\end{minipage}
\caption{Total solution time for different implementations and a range
  of block-solver tolerances $\epsilon$ for the Poisson problem (left) and
  the diffusion problem with spatially varying coefficients (right).\added{ cf.~\cite{BASTIAN2019417}.}}
\label{fig:solutiontime_diffusion}
\end{figure}

For solving the blocks related to $V_T^p$, two approaches have been implemented. In the first (named \textit{partially
matrix-free}), these
diagonal blocks are factorized using LAPACK and each iteration uses a backsolve. In the second approach
the diagonal blocks are solved iteratively to low accuracy using matrix-free sum factorization.
Both variants can be used in additive and multiplicative fashion.
Figure \ref{fig:solutiontime_diffusion}
shows that the partially matrix-free variant \replaced{is optimal for polynomial degree $p \leq 5$, but starting from $p = 6$,
  the fully matrix-free version starts to outperform all other options.}{is almost always the preferred one.}


\begin{figure}
\begin{center}
    \includegraphics[width=0.45\linewidth]{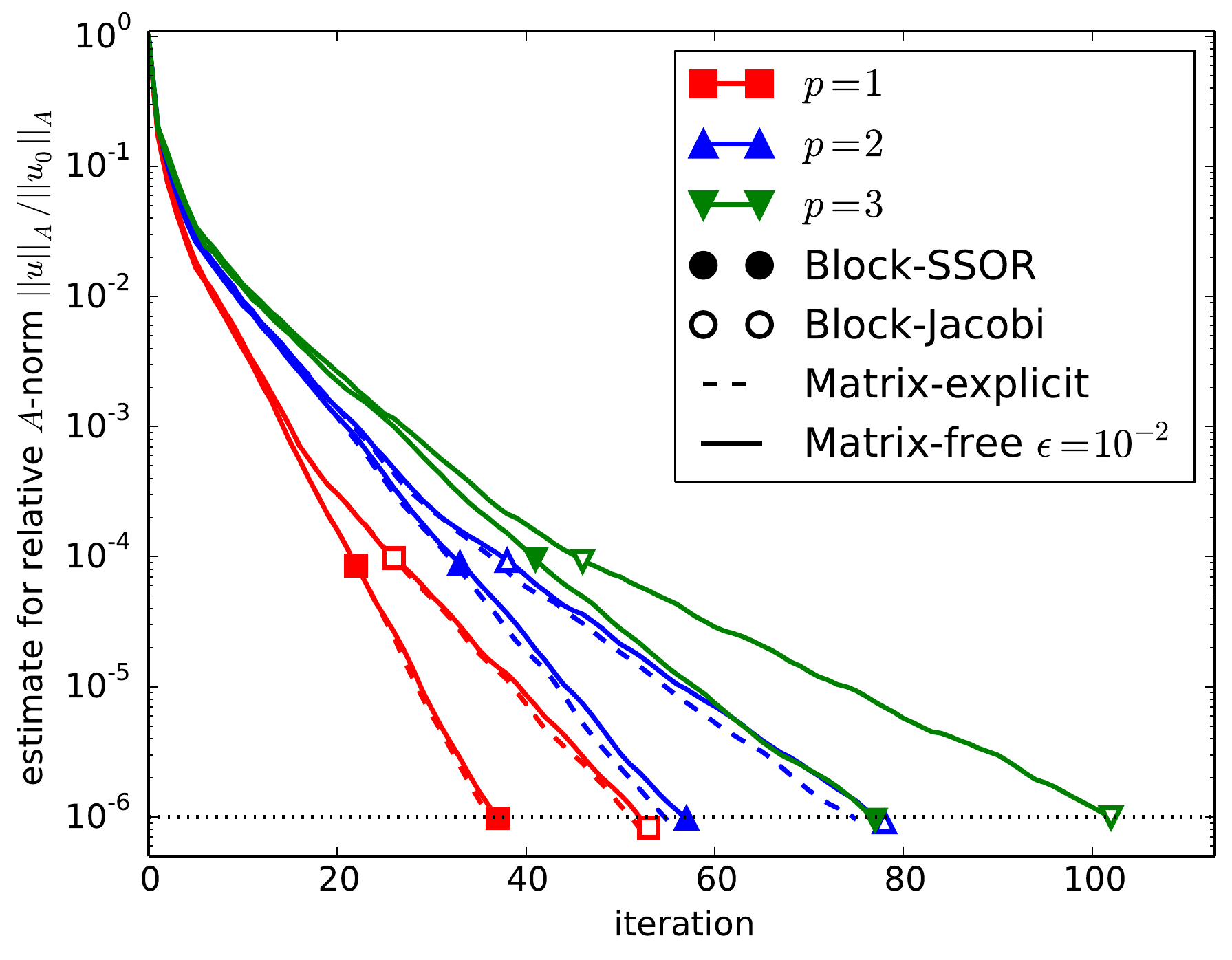}
\end{center}
  \caption{Convergence history for SPE10 benchmark. The relative
    energy norm is
    shown for polynomial degrees 1 (red squares), 2 (blue upward
    triangles) and 3 (green downward triangles). Results for the
    block-SSOR smoother are marked by filled symbols and results for
    the block-Jacobi smoother by empty symbols.\added{ cf.~\cite{BASTIAN2019417}.}}
  \label{fig:convergence_spe10}
\end{figure}

In order to demonstrate the robustness of our hybrid AMG-DG method we use the permeability
field of the SPE10 benchmark problem \cite{Christie2001} within a heterogeneous elliptic problem.
This is considered to be a hard test problem in the porous media community. The DG method
from \cite{Ern01042009} is employed. Figure \ref{fig:convergence_spe10} depicts results
for different variants and polynomial degrees run in parallel on 20 cores. A moderate increase with the polynomial degree
can be observed. With respect to time-to-solution (not reported) the additive (block Jacobi) partially matrix-free variant
is to be preferred for polynomial degree larger than one.

\subsection{Horizontal vectorization of Block Krylov methods}
  Methods like Multiscale FEM (see \replaced{S}{s}ection \ref{sec_ms}), optimization
  and inverse problems need to invert the same operator for many
  right-hand-side vectors.
  This leads to a block problem, by the following conceptual reformulation:
  \begin{equation*}
    \text{foreach}\quad i \in [0,N]: \text{solve } A x_i = b_i \qquad\rightarrow\qquad \text{solve } A X = B,
  \end{equation*}
  with matrices $X=(x_0,..x_N)$, $B=(b_0,..b_N)$. Such problems can be
  solved using Block Krylov solvers. The benefit is that the
  approximation space can grow faster, as the solver orthogonalizes
  the updates for all right-hand-sides. Even for a single
  right-hand-side Block Krylov based enriched Krylov methods can be
  used to accelerate the solution process.

  Preconditioners and the actual Krylov solver can be sped up using
  horizontal vectorization. Assuming $k$ right-hand-sides we observe
  that the scalar product yields a $k\times k$ dense matrix and has
  $O(k^2)$ complexity. While the mentioned larger approximation space
  should improve the convergence rate, this is only true for weaker
  preconditioners, therefore we pursued a different strategy and
  approximate the scalar product matrix by a sparse matrix, so that we
  again retain $O(k)$ complexity. In particular we consider the
  case of a diagonal or block-diagonal matrix. The diagonal matrix
  basically results in $k$ independent solvers running in parallel, so
  that the performance gain is solely based on SIMD vectorization and
  the associated favorable memory layout.

    \begin{figure}[htb]
    \centering
    \includegraphics[height=0.37\linewidth]{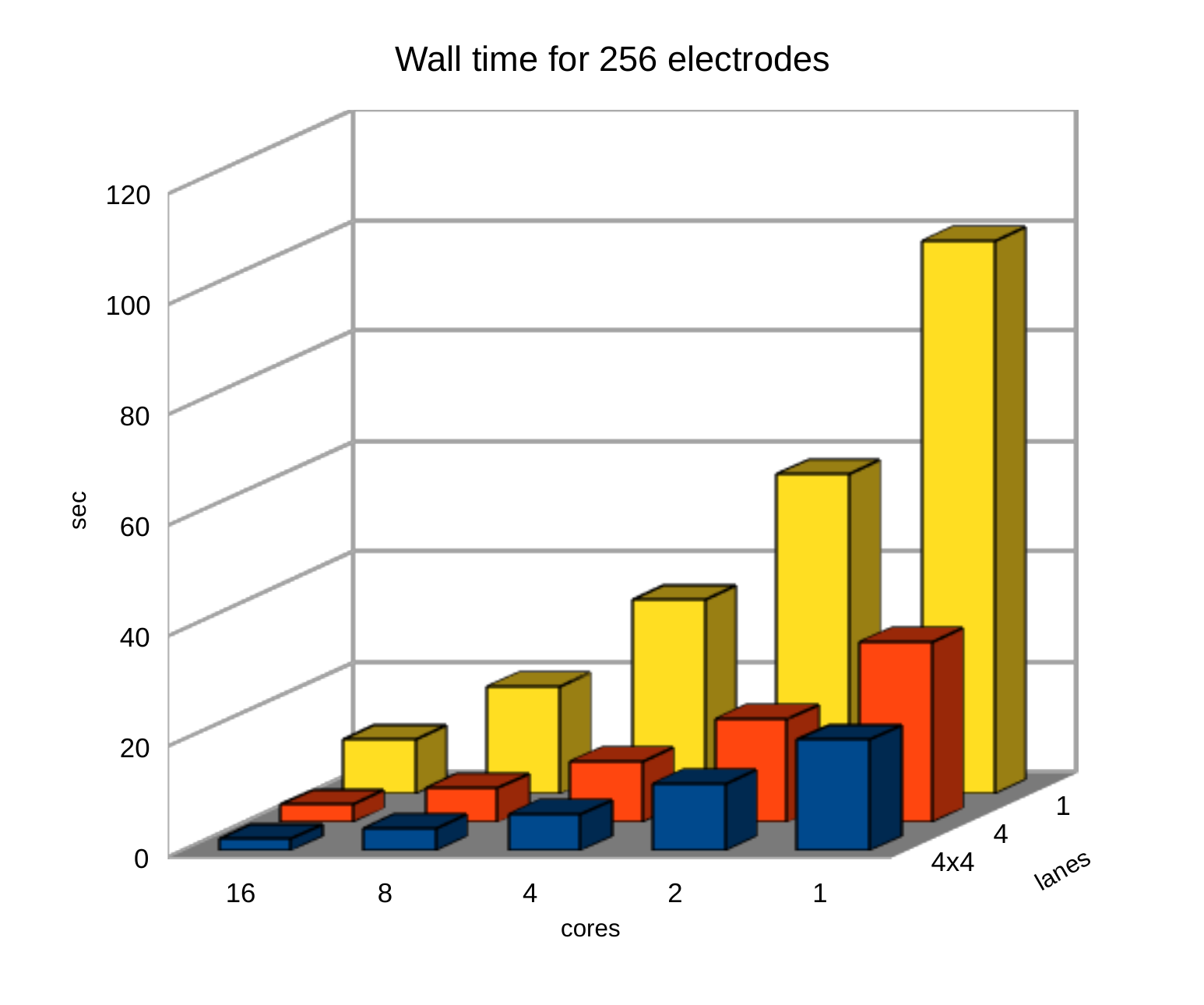}\qquad
    \includegraphics[height=0.37\linewidth]{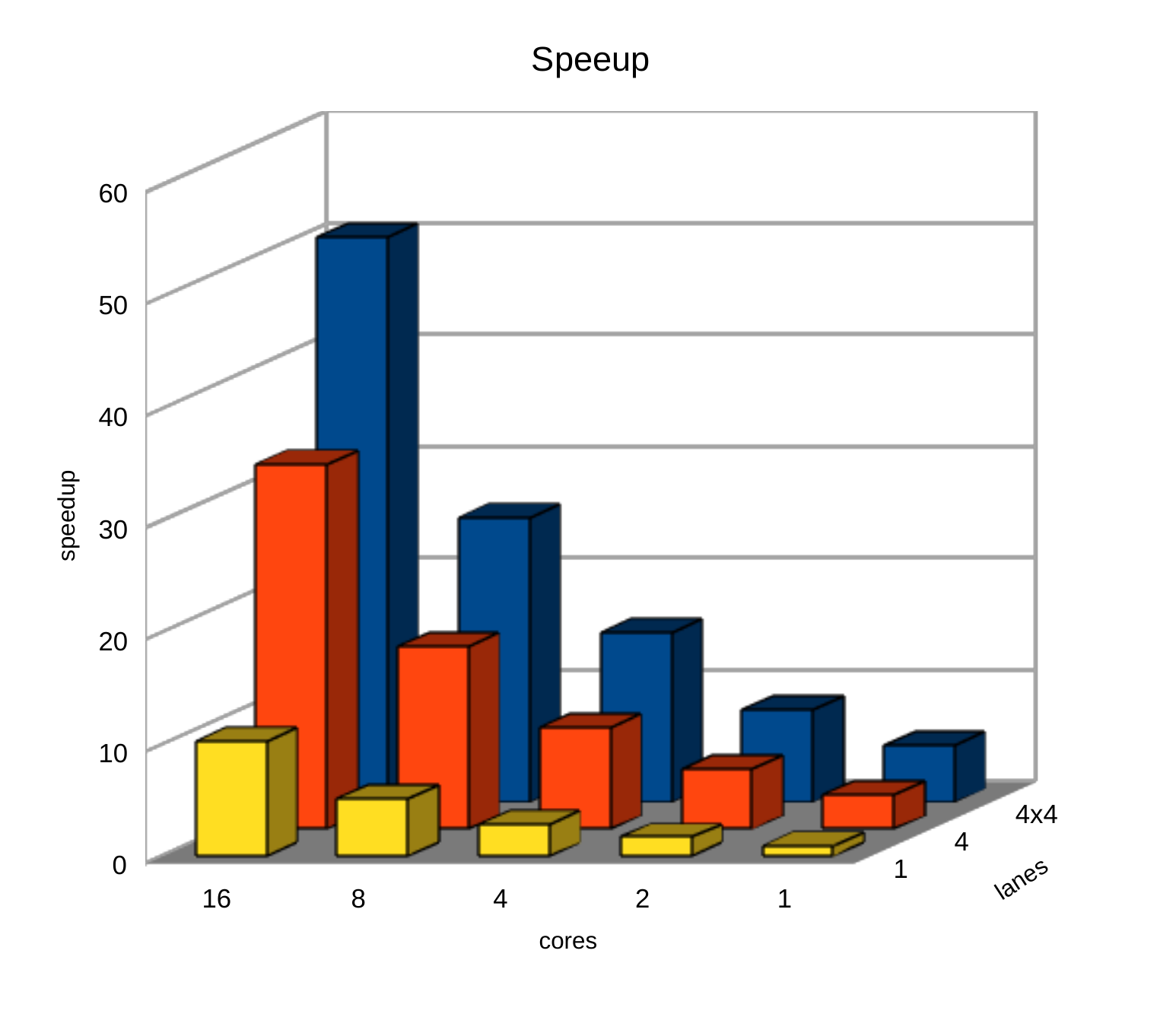}
    \caption{Horizontal vectorization of a linear solver for 256
      right-hand-side vectors. Timings on a Haswell-EP
      (E5-2698v3, 16 cores, AVX2, 4 lanes). Comparison with 1-16 cores
      and no SIMD, AVX (4 lanes), AVX ($4\times 4$ lanes).}
    \label{istl-simd}
  \end{figure}

  For the implementation in \istl we use platform portable C++ abstractions of
  SIMD intrinsics, building on the VC library\cite{kretz2012vc} and some
  \dune{} specific extentions. We use this to exchange the
  underlying data type of the right-hand-side and the solution vector,
  so that we no longer store scalars, but SIMD vectors.  This is possible
  when using generic programming techniques, like C++ templates, and
  yields a row-wise storage of the dense matrices $X$ and $B$. This
  row-wise storage is optimal and ensures a high arithmetic intensity.
  The implementations of the Krylov solvers have
  to be adapted to the SIMD data types, since some operations, like casts
  and branches, are not available generically for SIMD data types. As a side effect, all
  preconditioners, including the AMG, are now fully vectorized.

  Performance tests using 256 right-hand-side vectors for a 3D Poisson
  problem show nearly optimal speedup on a 64 core system (see
  \replaced{F}{f}igure~\ref{istl-simd}). The tests are carried out on a Haswell-EP
  (E5-2698v3, 16 cores, AVX2, 4 lanes). We observe a speedup of 50,
  while the theoretical speedup is 64.


\section{Adaptive Multiscale Methods}
\label{sec_ms}

\newcommand{\bdg}{{b}_{h}}
\newcommand{\rhs}{l_{h}}
\newcommand{\n}[0]{\bf n}
\newcommand{\faces}{\mathcal{F}}
\newcommand{\face}{e}
\newcommand{\Face}{E}
\newcommand{\tildefaces}{\tilde{\mathcal{F}}_h}
\newcommand{\innerfaces}[1]{\overcirc{\mathcal{F}}_#1}
\newcommand{\boundaryfaces}[1]{\overline{\mathcal{F}}_#1}
\newcommand{\Grid}{\mathcal{T}_H}
\newcommand{\grid}{\tau_h}

The main goal in the second funding phase was a distributed adaptive multilevel implementation of the
localized reduced basis multi-scale method (LRBMS~\cite{mso:lrbms2015}).
Like Multiscale FEM (MsFEM), \replaced{LRBMS is designed to work on heterogenous multiscale
or large scale problems. It performs particularly well for problems that exhibit scale separation}{LRBMS is designed to work on problems that exhibit scale separation}
with effects on both a fine and a coarse scale contributing to the global
behavior. Unlike MsFEM, LRBMS is best applied in multi-query settings
in which a parameterized PDE needs to be solved many times for different parameters.
As an amalgam of domain decomposition and model order reduction techniques,
the computational domain is partitioned into a coarse grid with each
macroscopic grid cell representing a subdomain for which, in an offline pre-compute
stage, local reduced bases are constructed.
Appropriate coupling \replaced{is then}{then is} applied
to produce a global solution approximation from localized data.
For increased approximation fidelity we can integrate localized global solution snapshots into the bases,
or the local bases can adaptively be enriched in the
online stage, controlled by a localized a-posteriori error estimator.

\subsection{Continuous problem and discretization}

We consider elliptic parametric multi-scale problems on a domain $\Omega \subset \mathbb{R}^d$
where we look for $p(\boldsymbol{\mu})\in Q$ that satisfy
\begin{align}\label{mso:para_problem}
  b(p(\boldsymbol{\mu}),q;\boldsymbol{\mu}) = l(q) \quad\quad\quad \textnormal{for all}\,q\in H^1_0(\Omega),
\end{align}
$\boldsymbol{\mu}$ are parameters with
$\boldsymbol{\mu} \in \mathcal{P}\subset \mathbb{R}^p$, $p\in\mathbb{N}$. We let
$\epsilon > 0$ be the multi-scale parameter associated with the fine scale.
\deleted{
For the discretization we first require a simplicial triangulation
$\mathcal{T}_H$ of $\Omega$ for the macro level. We call the elements
$T \in \mathcal{T}_H$ subdomains. We then require each subdomain be covered
with a partition $\tau_h(T)$ in a way that $\mathcal{T}_H$ and $\tau_h(T)$ are nested,
with $\bigcup_{\tau_h(T)} t = T$ and $\tau_h = \Sigma_{T \in \mathcal{T}_H} \tau_h(T)$
and the sets of faces of the coarse triangulation $\mathcal{F}_H$ and the fine
triangulation $\mathcal{F}_h$.
Next we define the broken (local) Sobolev spaces $H^1(\tau_h) \subset L^2(\Omega)$ as
$H^1(\tau_h) := \{ q\in L^2(\Omega) \ \vert\ q\vert_t \in H^1(t) \,\forall t \in \tau_h \}$,
and $H^1(\tau_h(T)) \subset L^2(T)$ accordingly.
}
For demonstration purposes we \deleted{now} consider a particular linear elliptic problem
setup in $\Omega\subset \mathbb{R}^d$ ($d=2,3$) that exhibits a multiplicative splitting in
the quantities affected by the multi-scale parameter $\epsilon$. \replaced{It is a model
for the so called global pressure $p(\boldsymbol{\mu}) \in H^1_0(\Omega)$ in
two phase flow in porous media, where the total scalar mobility $\lambda(\boldsymbol{\mu})$
is parameterized. $\kappa_\epsilon$ denotes the heterogenous permeability tensor
and $f$ the external forces. Hence, we seek $p$ that satisfies weakly in $H^1_0(\Omega)$,}
{This models
finding a pressure $p(\boldsymbol{\mu}) \in H^1_0(\Omega)$ with $\lambda(\boldsymbol{\mu})$
the total scalar mobility, external forces $f$ and permeability tensor $\kappa_\epsilon$
such that $p$ satisfies, weakly in $H^1_0(\Omega)$,
}
\begin{align}
  -\nabla \cdot (\lambda(\boldsymbol{\mu})\kappa_{\epsilon}\nabla p(\boldsymbol{\mu})) = f \quad\quad \textnormal{in}\  \Omega.
\end{align}
\replaced{With $A(x;\boldsymbol{\mu}) := \lambda(\boldsymbol{\mu})\kappa_{\epsilon}(x)$ this gives rise to the following definition of the forms in (\ref{mso:para_problem})}{
With these definitions in place we can write the continuous problem as
the (local) parametric bilinear forms}
\begin{eqnarray*}
  b(p(\boldsymbol{\mu}),q;\boldsymbol{\mu}) := \int_\Omega A(\boldsymbol{\mu}) \nabla p \cdot \nabla q, \quad
  l(q) :=  \int_{\Omega}{f q}.
\end{eqnarray*}

\added{For the discretization we first require a triangulation
$\mathcal{T}_H$ of $\Omega$ for the macro level. We call the elements
$T \in \mathcal{T}_H$ subdomains. We then require each subdomain be covered
with a fine partition $\tau_h(T)$ in a way that $\mathcal{T}_H$ and
$\tau_h := \Sigma_{T \in \mathcal{T}_H} \tau_h(T)$ are nested.
We denote by  $\mathcal{F}_H$ the faces of the coarse triangulation and
by  $\mathcal{F}_h$ the faces of the fine triangulation.}

\added{Let $V(\grid) \subset H^2(\grid)$ denote any approximate subset of the broken Sobolev space
$H^2(\tau_h) := \{ q\in L^2(\Omega) \ \vert\ q\vert_t \in H^2(t) \,\forall t \in \tau_h \}$.
We call $p_h(\boldsymbol{\mu}) \in V(\grid)$ an approximate solution of (\ref{mso:para_problem}), if}
\begin{align}\label{eq:broken}
  \bdg\big(p_h(\boldsymbol{\mu}),v;\boldsymbol{\mu}\big) &= \rhs(v; \boldsymbol{\mu}) &&\text{for all }v \in V(\grid).
\end{align}
\added{Here, the DG bilinear form $\bdg$ and the right hand side $\rhs$ are
chosen according to the SWIPDG method \cite{mso:ern10}, i.e.}
\begin{align*}
    \bdg(v,w;\boldsymbol{\mu}) &:= \sum_{t\in\grid}\int_t A(\boldsymbol{\mu})  \nabla v\cdot\nabla w
    + \sum_{e\in\faces(\grid)} \bdg^e(v, w; \boldsymbol{\mu})\\
    \rhs(v; \boldsymbol{\mu}) &:= \sum_{t\in\grid}\int_t fv,
   \intertext{where the DG coupling bilinear forms $\bdg^e$ for a face $e$ is given by}
   \bdg^e(v, w; \boldsymbol{\mu}) &:= \int_e \big< A(\boldsymbol{\mu}) \nabla v\cdot {\n_e}\big>[w]
    + \big< A(\boldsymbol{\mu}) \nabla w\cdot {\n_e} \big>[v]
    + \frac{\sigma_\face(\boldsymbol{\boldsymbol{\mu}})}{|e|^{\beta}} [v][w].
\end{align*}

\added{The LRBMS method allows for a variety of discretizations, i.e.
approximation spaces $V(\grid)$.
As a particular choice of an underlying high dimensional approximation space
we choose
$V(\grid) = Q^{k}_h := \bigoplus_{T \in \Grid} Q^{k,T}_h $, where the discontinuous local spaces are defined as}
\begin{align*}
  Q^{k,T}_h := Q^{k,T}_h(\tau_h(T)):=\{ q\in L^2(T) \ \vert\  q\vert_t \in \mathbb{P}_k(t) \,\forall t \in \tau_h(T) \}.
\end{align*}

\deleted{
The LRBMS method allows for a variety of discretizations of the local forms
to be used. We choose the local ansatz spaces $Q^{k,T}_h=Q^{k,T}_h(\tau_h(T))$
to be discontinuous,}

\deleted{and discretize $b^T_h$ with the SWIPDG method \cite{mso:ern10}.
Next we introduce the global coupling of these local approximations
along the faces $E \in \mathcal{F}_H$ of the coarse triangulation, again using
the SWIPDG method to arrive at the discretized bilinear form}

\deleted{for all $E\in\mathcal{F}_H$, all $\boldsymbol{\mu}\in\mathcal{P}$ and all
$p,q \in H^1(\tau_h)$,
with local coupling bilinear forms $  b^E_h $, the detailed defintion of which
can be found in \cite[Sec.~2.3]{mso:lrbms2015}.
This formulation naturally leads to solving a sparse blocked linear system where
the on-diagonal blocks represent the discretized $b^T_h(p(\boldsymbol{\mu}),q;\boldsymbol{\mu})$
and off-diagonal blocks contain the coupling matrices resulting from
discretizing $b^E_h(p(\boldsymbol{\mu}),q;\boldsymbol{\mu})$.
We call solutions of this global system $p_h(\boldsymbol{\mu})$.
}

\subsection{Model Reduction}

\replaced{For model order reduction in the LRBMS method we choose the reduced space
$Q_\text{red} := \bigoplus_{T \in \Grid}  Q^T_\text{red} \subset Q^{k}_h$ with local reduced approximation spaces
$Q^T_\text{red} \subset  Q^{k,T}_h$.
We denote $p_\text{red}(\boldsymbol{\mu})$ to be the reduced solution of (\ref{eq:broken})
in $Q_\text{red}$. This formulation naturally leads to solving a sparse blocked linear system similar
to a DG approximation with high polynomial degree on the coarse subdomain grid.}{
Model reduction techniques aim at finding approximate solutions in a reduced space
  $Q_\text{red}\subset Q^k_h$ that is of significantly smaller dimension
  $\dim Q_\text{red} \ll \dim Q^k_h $. One idea is to span $Q_\text{red}$ with solution
  snapshots to cover more of the solution manifold with fewer basis vectors.
  Manifesting the overall domain decomposition strategy, we construct local
  reduced spaces $Q^T_\text{red}\subset Q^{k,T}_h$ to define the coarse reduced space}
  \deleted{
  The natural conclusion is to then also search for a reduced solution
  $p_\text{red}\rightarrow Q_\text{red}(\mathcal{T}_H)$ for a parameter $\boldsymbol{\mu}\in\mathcal{P}$ that satisfies}

  The construction of subdomain reduced spaces $Q^T_\text{red}$ is again very flexible. Initialization
  with shape functions on $T$ up to order $k$ ensures a minimum fidelity. Basis extensions can be driven
  by a discrete weak greedy approach which incorporates localized solutions of the global system.
  Depending on available computational resources, and given a suitable localizable a-posteriori error estimator $\eta(p_\text{red}(\boldsymbol{\mu}), \boldsymbol{\mu})$,
   we can forego computing global high-dimensional solutions altogether
  and only rely on online enrichment to extend $Q^T_\text{red}$ \lq on the fly\rq. With online enrichment, given a reduced solution
  $p_\text{red}(\boldsymbol{\mu})$ for some arbitrary $\boldsymbol{\mu} \in \mathcal{P}$, we first compute local error indicators
  $\eta^T(p_\text{red}(\boldsymbol{\mu}), \boldsymbol{\mu})$ for all $T \in \mathcal{T}_H$. If
  $\eta^T(p_\text{red}(\boldsymbol{\mu}), \boldsymbol{\mu})$ is greater than some prescribed bound $\delta_\text{tol} > 0$
  we solve on a overlay region $\mathcal{N}(T) \supset T$ and extend $Q^T_\text{red}$ with
  $p_{\mathcal{N}(T)}(\boldsymbol{\mu}) \vert_{T}$. \added{Inspired by results in \cite{mso:henning} we set the overlay
  region's diameter $\diam(\mathcal{N}(T))$ of the order $\mathcal{O}(\diam(T)\vert log(\diam(T))\vert)$. In practice
  we use the completely on-line/off-line decomposable error estimator developed in \cite[Sec.~4]{mso:lrbms2015} which
  in turn is based on the idea of producing a conforming reconstruction of the diffusive flux
  $\lambda(\boldsymbol{\mu})\kappa_{\epsilon}\nabla_h p_h(\boldsymbol{\mu})$ in some Raviart-Thomas-Nédélec space
  $V^l_h(\tau_h) \subset H_{div}(\Omega)$ presented in \cite{Ern01042009,mso:vohralik}.}

  This process is then repeated until either a maximum
  number of enrichment steps occur or $\eta^T(p_\text{red}(\boldsymbol{\mu}), \boldsymbol{\mu})\leq\delta_\text{tol}$.

\begin{algorithm}[h]
      \label{mso:lrbms_alg}
      \begin{algorithmic}[1]
        \Require $P_\text{train} \subset \mathcal{P}$
        \Require Reconstruction operator $R_h(p_\text{red}(\boldsymbol{\mu})): Q_\text{red}(\mathcal{T}_H) \rightarrow   Q^k_{h}(\tau_h)$
        \Function{GreedyBasisGeneration}{$\delta_{grdy}$, $\eta(p_\text{red}(\boldsymbol{\mu}), \boldsymbol{\mu})$=None }

          \If{$\eta(p_\text{red}(\boldsymbol{\mu}), \boldsymbol{\mu})$ is not None}
            \State $E \gets \{ \eta(p_\text{red}(\mu_i), \mu_i ) \,\vert\, \mu_i \in \mathcal{P}_\text{train} \}$
          \Else
            \State $E \gets
              \{ \vert\vert R_h(p_\text{red}(\mu_i)) - p_h(\mu_i)\vert\vert \,\vert\, \mu_i \in \mathcal{P}_\text{train} \}$
          \EndIf
          \While{$E\neq\emptyset$ AND $max(E)\geq\delta_{grdy}$}
            \State $i \gets \textnormal{argmax}(E) $
            \State compute $p_h(\boldsymbol{\mu_i})$
            \ForAll{$T \in \mathcal{T}_H$}
              \State extend $Q_\text{red}^{T}$ with $p_h(\boldsymbol{\mu})\vert_T$
            \EndFor
            \State $E \gets E  \setminus E_i $
          \EndWhile
        \EndFunction
        \Statex
        \State Generate $\mathcal{T}_H$\Comment{Offline Phase}
        \ForAll{$T \in \mathcal{T}_H$}
          \State create $\tau_h(T)$
          \State init $Q_\text{red}^{T}$ with DG shape functions of order $k$
        \EndFor

        \State \textsc{GreedyBasisGeneration}($\cdots$) \Comment{Optional}

        \State compute $p_\text{red}(\boldsymbol{\mu})$ for arbitrary$\boldsymbol{\mu}$\Comment{Online phase}
        \ForAll{$T \in \mathcal{T}_H$} \Comment{Optional Adaptive Enrichment}
          \State  $\eta \gets \eta^T(p_\text{red}(\boldsymbol{\mu}), \boldsymbol{\mu})$
          \While{$\eta \geq \delta_\text{tol} $}
            \State compute $p_{\mathcal{N}(T)}(\boldsymbol{\mu})$
            \State $Q_\text{red}^{T} \gets p_{\mathcal{N}(T)}(\boldsymbol{\mu}) \vert_{T}$
          \EndWhile
        \EndFor
      \end{algorithmic}
      \caption{Schematic representation of the LRBMS pipeline.}
\end{algorithm}

\subsection{Implementation}

We base our MPI-parallel implementation of LRBMS on the serial version developed
previously. In this setup the high-dimensional quantities and all grid structures
are implemented in \dune{}. The model order reduction as such is implemented
in Python using pyMOR \cite{mso:pymor2016}.
The model reduction algorithms in pyMOR follow a solver agnostic design principle. Abstract interfaces
allow for example projections, greedy algorithms or reduced data reconstruction to be written
without knowing details of the PDE solver backend.
The global macro grid $\mathcal{T}_H$
can be any MPI-enabled \dune{} grid manager with adjustable overlap size for the
domain decomposition, we currently use \dunemodule{YaspGrid}. The fine grids $\tau_h(T)$ are constructed
using the same grid manager as on the macro scale, with \replaced{MPI subcommunicators}{rank-local MPI communicators}.\added{These are currently
limited to a size of one (rank-local), however the overall scalability could benefit from dynamically
sizing these subcommunicators to balance communication overhead and computational intensity as demonstrated
in \cite[Sec.~2.2]{mso:exasteel}.}
The assembly of the local (coupling) bilinear forms is done in \dunemodule{GDT} \cite{mso:gdt},
with pyMOR/Python bindings facilitated through \dunemodule{XT} \cite{mso:xt}, where
\dunemodule{Grid-Glue} \cite{mso:glue2016} generates necessary grid views for
the SWIPDG coupling between otherwise unrelated grids. Switching to \dunemodule{Grid-Glue} constitutes
a major step forward in robustness of the overall algorithm, compared
to our previous manually implemented approach to matching independent local grids for
coupling matrices assembly.

\begin{figure}[htb]
  \centering
  \scalebox{0.45}{\input{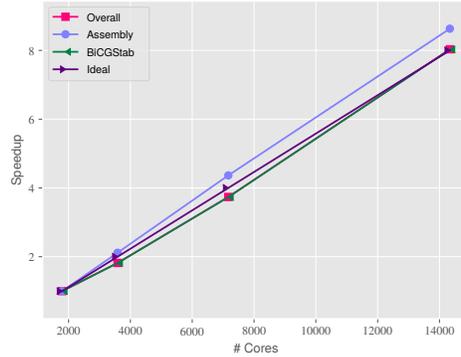}}
  \caption{Localized Reduced Basis Method: Block-SWIPDG speedup results; linear system solve (green),
    discretization and system assembly (blue), theoretic ideal speedup (violet)
    and actual achieved speedup for the overall run time (red).
    Simulation on $\sim7.9 \cdot 10^6 $ cubical cells shows minimum $94\% $
    parallel efficiency, scaling from $64$ to $512$ nodes (SuperMUC Phase 2).}
   \label{fig:blockswipd}
\end{figure}

We have identified three major challenges in parallelizing all the steps in LRBMS:

\noindent
 \textbf{1. Global solutions $p_{h}(\boldsymbol{\mu})$ of the blocked system in Equation~\ref{mso:bilin_blocked} with
  an appropriate MPI-parallel iterative solver.}
  With the serial implementation already using \istl{} as the backend for matrix and vector data,
  we only had to generate an appropriate communication configuration for the blocked SWIPDG matrix structure to make
  the BiCGStab solver usable in our context.
  This setup we then tested on the SuperMUC Petascale System in Garching. The results in  Figure~\ref{fig:blockswipd}
  show very near ideal speedup from 64 nodes with 1792 MPI ranks up to a full island with 512 nodes and 14336 ranks.

\noindent
  \textbf{2. (Global) Reduced systems also need a distributed solver.}
  By design all reduced quantities in pyMOR are, at the basic, unabstracted level, NumPy arrays~\cite{mso:numpy}.
  Therefore we cannot easily re-use the \istl{} based solvers for the high-dimensional systems.
  Our current implementation gathers these reduced system matrices from all MPI-ranks to rank 0, recombines
  them, solves the system with a direct solver and scatters the solution.
  There is great potential in making this step more scalable by either using a distributed sparse direct solver like
  Mumps \cite{mso:mumps} or translating the data into the \istl{} backend.

\noindent
  \textbf{3. Adaptive online enrichment is inherently load imbalanced due to its localized error estimator guidance.}
  The load imbalance results from one rank idling while waiting to receive updates to a basis on a subdomain
  in its overlap region from another rank. This idle time can be minimized by encapsulating
  the update in a \lstinline{MPIFuture} described in Subsection~\ref{subsec::async_abstraction}. This will allow the rank
  to continue in its own enrichment process until the updated basis is actually needed in a subsequent step.

\section{Uncertainty Quantification}
\label{sec_uq}
The solution of stochastic partial differential equations (SPDEs) is characterized by extremely high dimensions and poses great (computational) challenges.
Multilevel Monte Carlo (MLMC) algorithms attract great interest due to their superiority over the standard Monte Carlo approach.
Based on Monte Carlo (MC), MLMC retains the properties of independent sampling.
To overcome the slow convergence of MC, where many computationally
expensive PDEs have to be solved, MLMC combines in a proper way cheap MC estimators and expensive MC estimators, achieving (much) faster convergence.
One of the critical components of the MLMC algorithms is the way in which the coarser levels are selected.
The exact definition of the levels is an open question and different approaches exist.
In the first funding phase, Multiscale FEM was used as a coarser level in MLMC. During the second phase, the developed parallel MLMC algorithms for uncertainty quantification were further enhanced. The main focus was on exploring the  capabilities of the renormalization approach for defining the coarser levels in the MLMC algorithm, and on using MLMC as a coarse grained parallelization approach.

Here, we employ MLMC to exemplarily compute the mean flux through saturated porous media with prescribed pressure drop
and known distribution of the random coefficients.

{\bf Mathematical problem. }
As a model problem in $\mathbb{R}^2$ or $\mathbb{R}^3$, we consider steady state single phase flow in random porous media:
\begin{equation*}
-\nabla \cdot [k(x, \omega) \nabla p(x, \omega)] = 0 \mbox{ for } x \in D = (0,1)^d, \omega \in \Omega
\end{equation*}
subject to the boundary conditions $p_{x=0} = 1$ and $p_{x=1} = 0$ and zero flux on other boundaries.
Here $p$ is pressure, $k$ is scalar permeability, and  $\omega$ is random vector.
The quantify of interest is the mean (expected value) $E[Q]$ of the total flux $Q$ through the inlet of the unit cube i.e., $Q(x, \omega) := \int_{x=0} k(x,\omega) \partial_n p(x,\omega) dx$.
Both the coefficient $k(x, \omega)$ and the solution $p(x, \omega)$ are subject to uncertainty, characterized by the random vector $\omega$ in a properly defined random space $\Omega$. For generating permeability fields we consider practical covariance
$C(x, y) = \sigma^{2} \text{exp}(- ||x - y||_2 / \lambda)$.
An algorithm based on forward and inverse Fourier transform over the circulant covariance matrix is used
to generate the permeability field. For solving the deterministic PDEs a Finite Volume method on a cell centered grid is used \cite{iliev2017renormalization}. More details and further references can be found in a previous paper \cite{mohring2015uncertainty}.

{\bf Monte Carlo simulations. }
To quantify the uncertainty, and compute the mean of the flux we use a MLMC algorithm. Let $\omega_{M}$ be a random vector over a properly defined probability space, and $Q_{M}$ be the corresponding flux. It is known that $E[Q_{M}]$ can be made arbitrary close to $E[Q]$ by choosing $M$ sufficiently large. The standard MC algorithm convergences very slowly, proportionally to the variance over the square root of the number of samples, which makes it often unfeasible.
MLMC introduces $L$ levels with the $L$-th level coinciding with the considered problem, and exploits the telescopic sum identity:
\begin{equation*}
E[Q_M^L(\omega)]=E[Q_M^0(\omega)] + E[Q_M^1(\omega)-Q_M^0(\omega)] + ...  E[Q_M^L(\omega)-Q_M^{L-1}(\omega)]
\end{equation*}
The notation $Y^l=Q^1-Q^{l-1}$ is also used.
The main idea of MLMC is to properly define levels, and combine a large number of cheap simulations, that are able to approximate the variance well, with a small number of expensive correction simulations providing the needed accuracy. For details on Monte Carlo and MLMC we refer to previous publications \cite{iliev2017renormalization, mohring2015uncertainty} and the references therein. Here, the target is to estimate the mean flux on a fine grid, and we define the levels as discretizations on coarser grids. In order to define the permeability at the coarser levels we use the renormalization approach.

MLMC has previously run the computations at each level with the same tolerance. However, in order to evaluate the number of samples needed per level, one has to know the variance at each level. Because the exact variance is not known in advance, MLMC starts by performing simulations with a prescribed, moderate number of samples per level. The results are used to evaluate the variance at each level, and thus to evaluate the number of samples needed per level. This procedure can be repeated several times in an Estimate--Solve cycle. At each estimation step, information from all levels is needed, which leads to a synchronization point in the parallel realization of the algorithm. This may require dynamic redistribution of the resources after each new evaluation.

MLMC can provide a coarse graining in the parallelization. A well balanced algorithm has to account for several factors: (i) How many processes should be allocated per level; (ii)  How many processes should be allocated per deterministic problem including permeability generation; (iii) How to parallelize the permeability generation; (iv) Which of the parallelization algorithm for deterministic problems available in \exadune{} should be used; (v) Should each level be parallelized separately and if not, how to group the levels for efficient parallelization. The last factor is the one giving coarse grain parallelization opportunities. For the generation of the permeability, we use the parallel MPI implementation of the FFTW library. As deterministic solver, we use a parallel implementation of the conjugate gradient scheme preconditioned with AMG, provided by \istl{}. Both of them have their own internal domain decomposition.

We shortly discuss one Estimate-Solve cycle of the MLMC algorithm. Without loss of generality we assume 3-level MLMC. Suppose that we have already computed the required number of samples per level (i.e., we are after Estimate and before Solve). Let us denote by $N_{i}, i = \{0, 1, 2\}$ the number of required realizations per level for $\widehat{Y_{l}}$, by $p_{i}$ the number of processes allocated per $\widehat{Y_{i}}$, by $p_{l_{i}}^{g}$ the respective group size of processes working on a single realization, by $n$ the number of realizations for each group of levels, with $t_{i}$ the respective time for solving a single problem once, and finally with $p^\text{total}$ the total number of available processes. Then we can compute the total CPU time for the current Estimate-Solve cycle as
\begin{equation*}
 T^\text{total}_\text{CPU} =  N_{0} t_{0} + N_{1} t_{1} + N_{2} t_{2}.
\end{equation*}
Ideally each process should take $ T^{p}_\text{CPU} = T^\text{total}_\text{CPU} / p^\text{total}$.
Dividing the CPU time needed for one  $\widehat{Y_{i}}$ by $T^{p}_\text{CPU}$, we get a continuous value for the number of processes on a given level $ p_{i}^\text{ideal} = N_{i} t_{i} / T^{p}_\text{CPU}$ for $i = \{0, 1, 2\}$.
Then we can take $ p_{i} = \left \lfloor{p_{i}^\text{ideal}}\right \rfloor$.
To obtain an integer value for the number of processes allocated per level, first we construct a set of all possible splits of the original problem as a combination of subproblems (e.g., parallelize level 2 separately and the combination of levels 0 and 1, or parallelize all levels simultaneously, etc.). Each element of this set is evaluated independently, and all combinations of upper and lower bounds are calculated, such that ${p_{i}^\text{ideal}}$ is divisible by $p_{l_{i}}^{g}$, $\sum_{l=0}^{2} p_{i} < p^\text{total}$ and $p_{i} \le N_{i} p_{l_{i}}^{g}$. Traversing,  computing and summing the computational time needed for each element gives us a time estimation. Then we select the element (grouping of levels) with minimal computational time. To tackle the distribution of the work on a single level, a similar approach can be employed. Due to the large dimension of the search tree a heuristic technique can be employed. Here we consider a simple predefined group size for each deterministic problem, having in mind that when using AMG the work for a single realization at the different levels is proportional to the unknowns at this level.

\begin{figure}[htb]
\begin{center}
\includegraphics[width=0.5\textwidth]{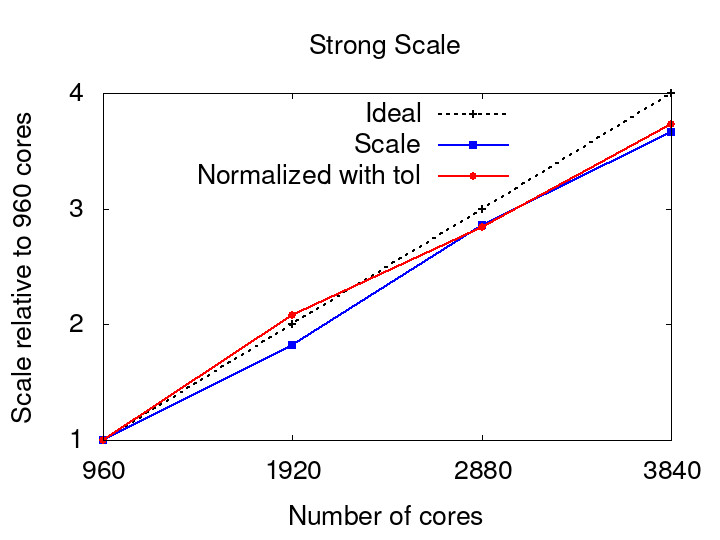}
\caption{Scalability of the MLMC approach}
\label{fig:2d_g11}
\end{center}
\end{figure}

{\bf Numerical experiments.} Results for a typical example are shown in \replaced{F}{f}igure~\ref{fig:2d_g11}. The parameters are $\sigma=2.75, \lambda=0.3$. The tests are done on SuperMUC-NG, LRZ Munich on a dual Intel Skylake Xeon Platinum 8174 node. Due to the stochasticity of the problem, we plot the speedup multiplied with the proportion of the tolerance.
The renormalization has shown to be a very good approach for defining the coarser levels in MLMC. The proposed parallelization algorithm gives promising scalability results. It is  weakly coupled to the number of samples that MLMC estimates. Although the search for an optimal solution is an NP-hard problem, the small number of levels enables a full traversing of the search tree. It can be further improved by automatically selecting the number of processes per group that solves a single problem. One also may consider introducing interrupts between the MPI communicators on a level to further improve the performance.

\section{Land-Surface Flow Application}
\label{sec_app}
To test some of the approaches developed in the \exadune{} project, especially the usage of sum-factorized operator evaluation with more complex problems, we developed an application
to simulate coupled surface-subsurface flow for larger geographical areas. This is a topic
with high relevance for a number of environmental questions from soil protection and groundwater
quality up to weather and climate prediction.

One of the central aims of the new approach developed in the project is to be able to relate a physical
meaning to the parameter functions used in each grid cell. This is not possible with the traditional
structured grid approaches as the necessary resolution would be prohibitive. To avoid the excessive
memory requirements of completely unstructured grids we build on
previous results for block-structured meshes and use a 2-dimensional unstructured grid on the
surface which is extruded in a structured way in vertical
direction. However, more flexible discretization
schemes are needed for such grids, compared to the usual cell-centred Finite-Volume approaches.

\subsection{Modelling and numerical approach}

To describe subsurface flow we use Richards equation \cite{richards-la31}
\begin{equation*} \label{eq:model-rich-mc}
\frac{\partial\theta(\psi)}{\partial t} - \nabla\cdot \left[k(\psi)\left(\nabla \psi + \mathbf{e}_g\right)\right] + q_w = 0
\end{equation*}
where $\theta$ is the volumetric water content, $\psi$ the soil water potential, $k$ the hydraulic conductivity, $\mathbf{e}_g$ the unit vector pointing in the direction of gravity and $q_w$ a volumetric source or sink term.

In nearly all existing models for coupled surface-subsurface flow, the kinematic-wave approximation is used for surface flow, which only considers surface slope as driving force and does not even provide a correct approximation of the steady-state solution. The physically more realistic shallow-water-equations are used rarely, as they are computationally expensive. We use the diffusive-wave approximation, which still retains the effects of water height on run-off, yields a realistic steady-state solution and is a realistic approximation for flow on vegetation covered ground \cite{alonso-2008}:
\begin{equation} \label{eq:model-dwa}
\frac{\partial h}{\partial t} - \nabla\cdot\left[D(h,\nabla h)\nabla(h+z)\right] = f_c,
\end{equation}
where $h$ is the height of water over the surface level $z$,
$f_c$ is a source-sink term (which is used for the coupling) and the diffusion coefficient $D$ is given by
\begin{equation*} \label{eq:model-dwa-diffusion-coef}
D(h,\nabla h) = \frac{h^\alpha}{C\cdot\|\nabla(h+z)\|^{1-\gamma}}
\end{equation*}
with $\|\cdot\|$ refering to the Euclidean norm
and
$\alpha$, $\gamma$ and $C$ are empirical constants. In the following we use $\alpha = \frac53$ and $\gamma = \frac12$ to obtain Manning's formula and a friction coefficient of $C=1$.

Both equations are discretised with a cell-centered Finite-Volume
scheme and alternatively with a SWIPDG scheme in space (see \replaced{S}{s}ection \ref{sec_solvers}) and an appropriate diagonally implicit Runge-Kutta scheme in time for the subsurface and an explicit Runge-Kutta scheme for the surface flow. Upwinding is used for the calculation of conductivity in subsurface flow \cite{bastian-2014} and for the water height in the diffusion term in surface flow.

First tests have shown that the formulation of the diffusive-wave approximation from the literature as given by equation \eqref{eq:model-dwa} does not result in a numerically stable solution if the gradient becomes very small, as then a gradient approaching zero is multiplied by a diffusion coefficient going to infinity. A much better behaviour is achieved by rewriting the equation as
\begin{equation*}
\frac{\partial h}{\partial t} - \nabla \cdot \left[\frac{h^\alpha}{C} \cdot  \frac{\nabla(h+z)}{\|\nabla(h+z)\|^{1-\gamma}}\right] = f_c,
\end{equation*}
where the rescaled gradient $\frac{\nabla(h+z)}{\|\nabla(h+z)\|^{1-\gamma}}$ is always going to zero when $\nabla(h+z)$ is going to zero as long as $\gamma < 1$ and the new diffusion coefficient $h^\alpha / C$ is bounded.

Due to the very different time-scales for surface and subsurface flow, an operator-splitting approach is used for the coupled system. A new coupling condition has been implemented, which is a kind of Dirichlet-Neumann coupling, but guarantees a mass-conservative solution. With a given height of water on the surface (from the initial condition or the last time step modified by precipitation and evaporation), subsurface flow is calculated with a kind of Signorini boundary condition, where all surface water is infiltrated in one time step as long as the necessary gradient is not larger than the pressure resulting from the water ponding on the surface (in infiltration) and potential evaporation rates are maintained as long as the pressure at the surface is not below a given minimal pressure (during evaporation). The advantage of the new approach is that it does not require a tracking of the sometimes complicated boundary between wet and dry surface elements, that it yields no unphysical results and that the solution is mass-conservative even if not iterated until convergence.

Parallelisation is obtained by an overlapping or non-overlapping domain-de\-com\-po\-si\-tion (depending on the grid).
However, only the two-dimensional surface grid is partitioned whereas the vertical direction is kept on one process due
to the strong coupling. Thus there is also no need for communication of surface water height for the coupling, as the
relevant data is always stored in the same process. The linear equation systems are solved with the BiCGstab-solver from
\istl{} with \added{Block-}ILU0 as preconditioner.
\added{The much larger mesh size in horizontal direction compared to the vertical direction results in strong coupling of the unknowns in the vertical direction. The Block-ILU0 scheme provides an almost exact solver of the strongly coupled blocks in the vertical direction and is thus a very effective scheme. Furthermore, one generally has a limited number of cells in the vertical direction and extends the mesh in horizontal direction to simulate larger regions. Thus the good properties of the solver are retained when scaling up the size of the system.}

\subsection{Performance Optimisations}

As the time steps in the explicit scheme for surface flow can get very small due to the stability limit, a significant speedup can be achieved by using a semi-implicit scheme, where the non-linear coefficients are calculated with the solution from the previous time step or iteration. However, if the surface is nearly completely dry, this could lead to numerical problems, thus an explicit scheme is still used under nearly dry conditions with an automatic switching between both.

While matrix-free DG solvers with sum-factorization can yield excellent per node performance (\replaced{S}{s}ection~\ref{subsec:matfreedg}), it is a rather tedious task to implement for new partial differential equations. \replaced{Therefore, a code-generation framework is currently being developed}{Therfore, a code-generation framework has been} in a project related to \exadune{} \cite{kempf-2019}. The framework is used to implement an alternative optimized version of the solver for Richards equation as this is the computationally most expensive part of the computations. Expensive material functions like the van Genuchten model including several power functions are replaced by cubic spline approximations, which allow a fast vectorized incorporation of flexible material functions to simulate strongly heterogeneous systems. Sum-factorisation is used in the operator evaluations for the DG-discretization with a selectable polynomial degree.

A special pillar grid has been developed as \replaced{a first realisation of a 2.5D grid \cite{kempf-2019}. It adds a vertical dimension to a two-dimensional grid (which is either structured or unstructured). However, as the current version still produces a full three-dimensional mesh at the moment, future developments are necessary to exploit the full possibilities of the approach.}{an efficient realisation of a 2.5D grid \cite{kempf-2019}. It adds a vertical dimension to a two-dimensional grid (which is either structured or unstructured) and optimally exploits known structural properties in the sum-factorised kernels.}

\subsection{Scalability and Performance Tests}

\begin{figure}[htb]
  \begin{center}
    \includegraphics[width=0.8\textwidth]{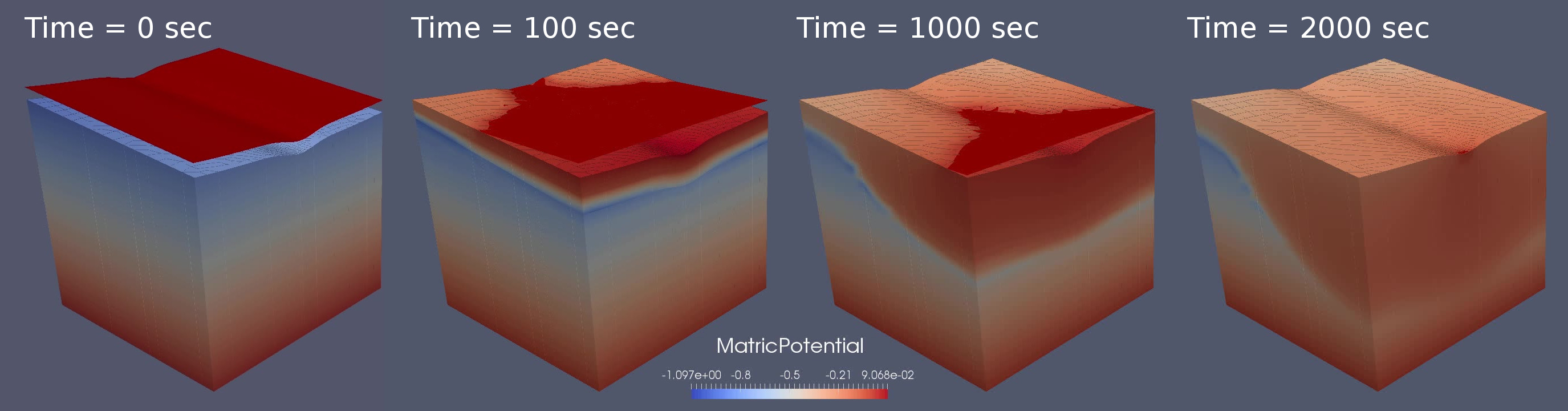}
    \includegraphics[width=0.8\textwidth]{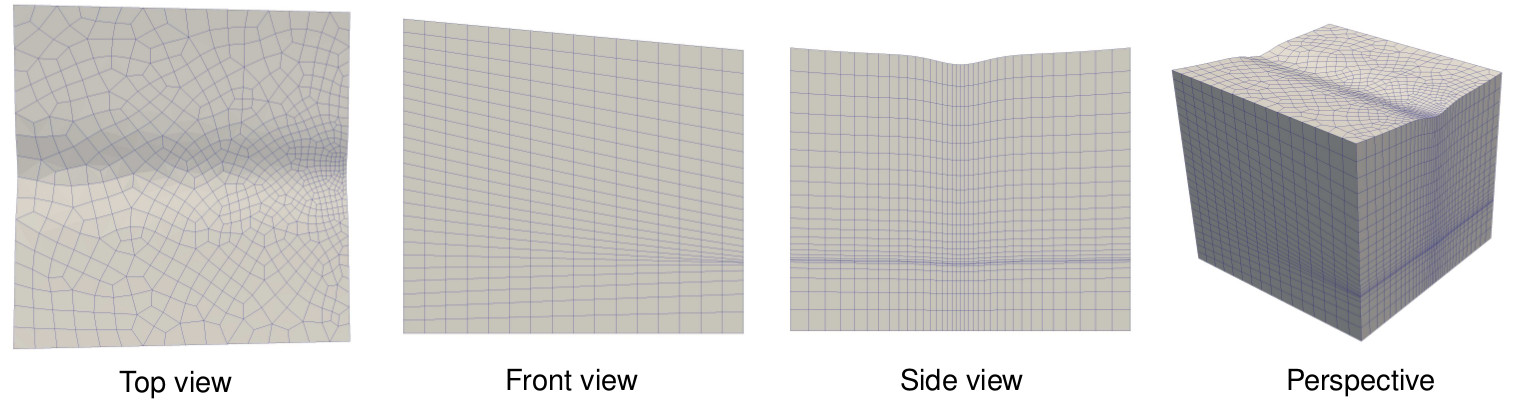}
  \end{center}
\caption{Surface runoff and infiltration of 5 cm water into a dry coarse sand (top) and the unstructured 2.5D mesh used for the simulations (bottom).}
\label{pic:infiltration}
\end{figure}

\begin{figure}[htb]
\begin{center}
    \includegraphics[width=0.48\textwidth]{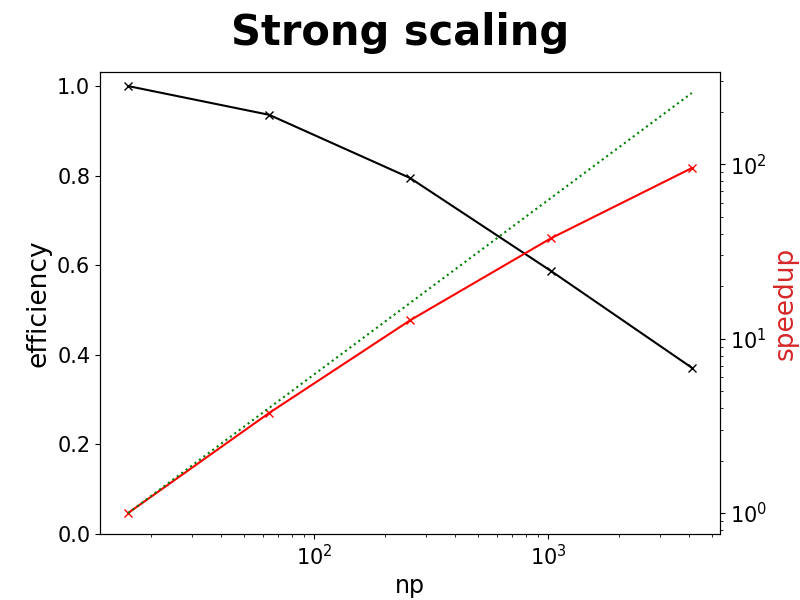}
    \includegraphics[width=0.48\textwidth]{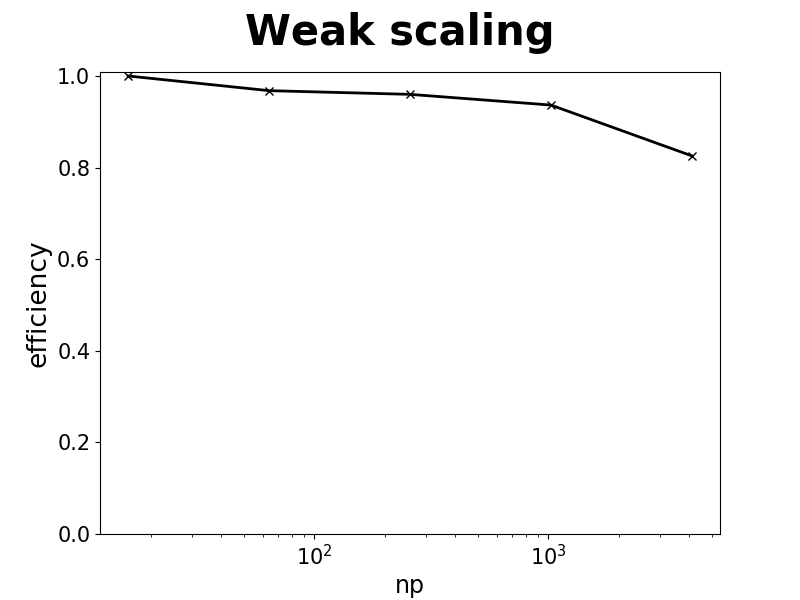}
\end{center}

\caption{\replaced{Results of a strong (left)  and weak (right) scalability test with a coupled run-off and infiltration experiment on 1 to 256 nodes (16 to 4096 Intel Xeon E5-2630 v3 2.4GHz~CPU cores) of the bwForCluster at IWR in Heidelberg.}{Domain decomposition and result for 5 time-steps of a coupled run-off and infiltration experiment (left) with $80 \times 80 \times 80$ elements and a polynomial degree of one. Very good strong scaling is achieved for up to 16 processes (right) on a dual Intel E5-2660 10-Core server.}}
\label{fig:strongweakscaling}
\end{figure}

\begin{figure}[htb]
  \begin{center}
    \includegraphics[width=\textwidth]{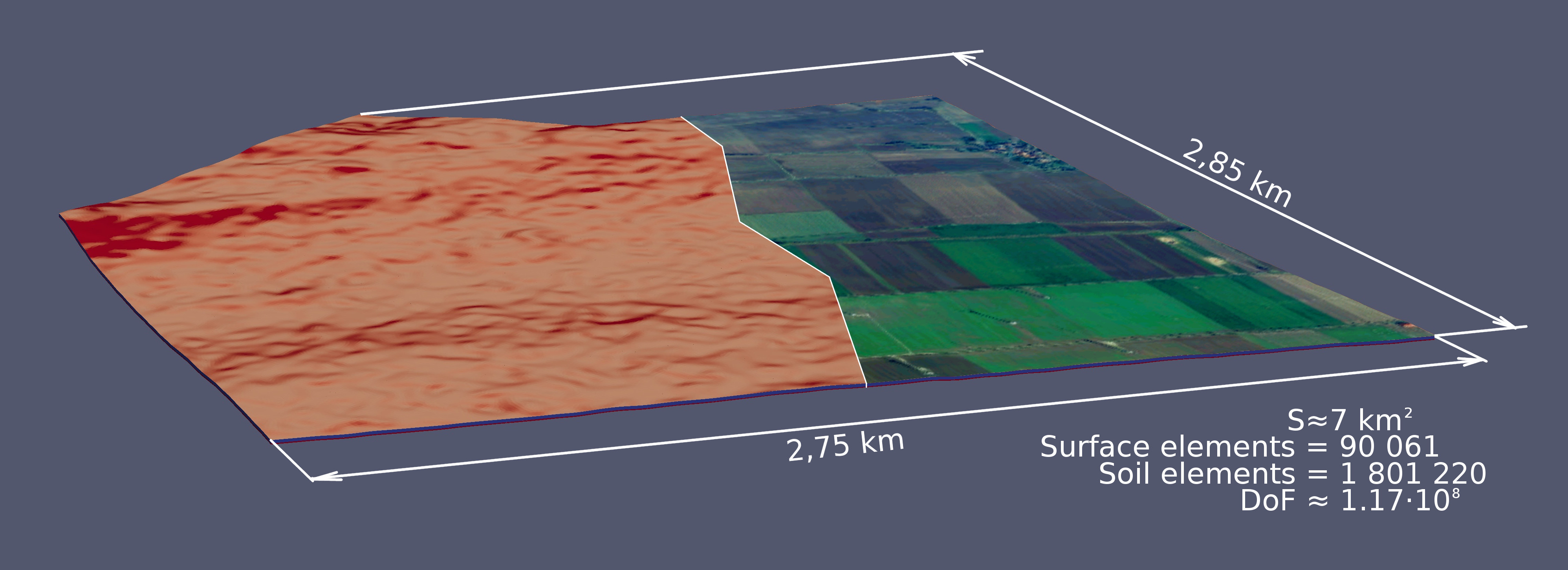}
  \end{center}
\caption{\added{Pressure distribution with an overlay of the landscape taken from Google Earth calculated in a simulation of water transport in a real landscape south of Brunswick simulated on 30 nodes with 1200 cores of HLRN-IV in Göttingen (2x Intel Skylake Gold 6148 2.4GHz~CPUs).}}
\label{pic:landscape}
\end{figure}

Extensive tests covering infiltration as well as exfiltration have been performed (e.g. Figure~\ref{pic:infiltration}) to test the implementation and the new coupling condition. \replaced{Good scalability is achieved in strong as well as in weak scaling experiments on up to 256 nodes and 4096 cores of the bwForCluster in Heidelberg (Figure~\ref{fig:strongweakscaling}). Simulations for a large region with topographical data taken from a digital elevation model (Figure~\ref{pic:landscape}) have been conducted as well.}{Good scalability is achieved in first strong (Figure~\ref{fig:strongscaling}) as well as in weak scaling experiments on a small computing cluster. In the latter there is a small increase in computation time from one to four processes, but thereafter we observe a constant parallel efficiency (not shown).}

\begin{figure}[htb]
\begin{center}
\includegraphics[width=0.6\textwidth]{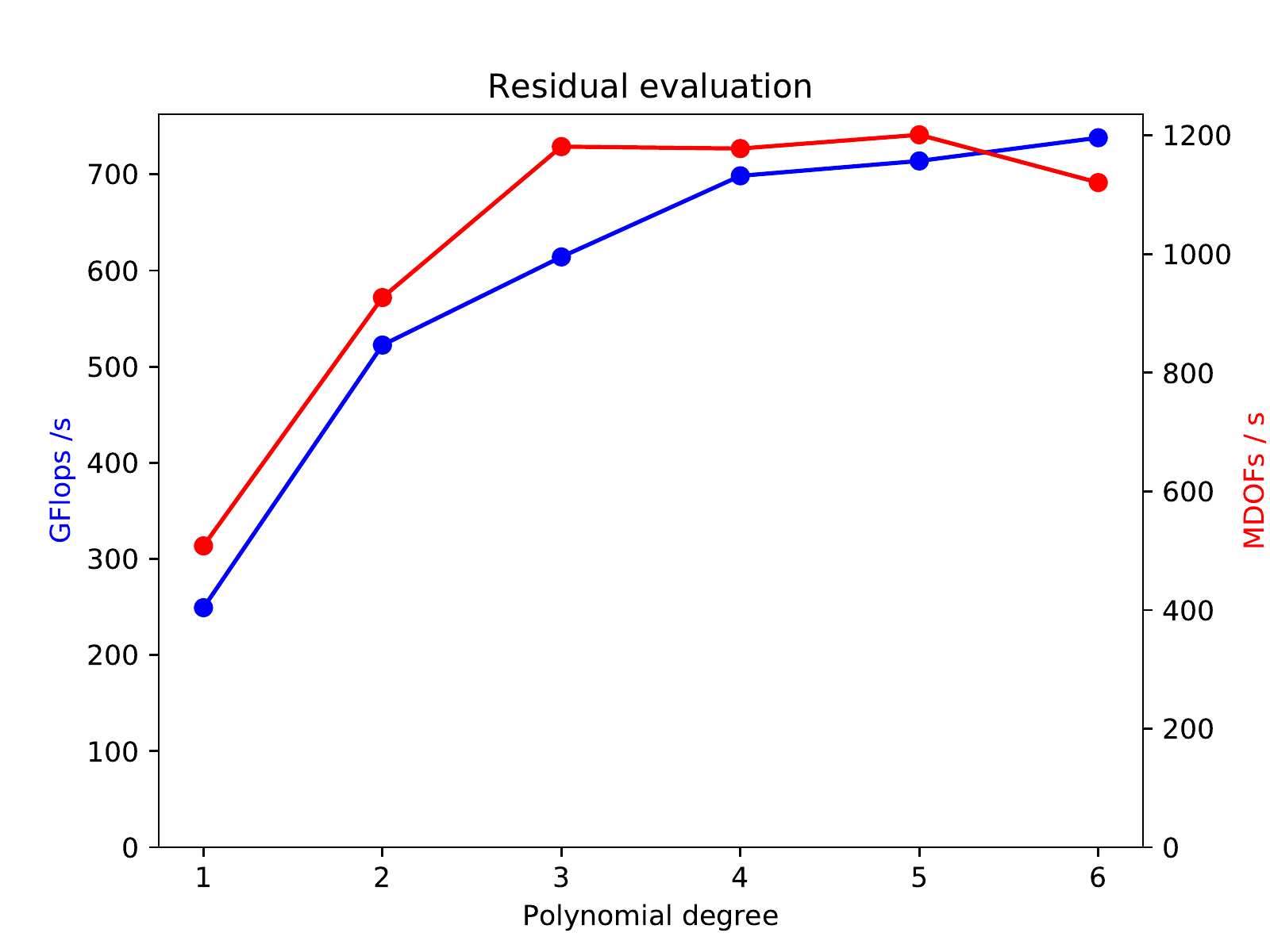}
\end{center}
\caption{Performance of the Richards' solver implemented with the code generation framework for \exadune{}}
\label{fig:performance}
\end{figure}

With the generated code-based solver for Richards equation a substantial fraction of the systems peak performance (up to 60 \% on a Haswell-CPU) can be utilized due to the combination of sum factorization and vectorisation (Figure~\ref{fig:performance}). For Richards equation (as for other PDEs before) the number of millions of degrees of freedom per second is independent of the polynomial degree with this approach.
\replaced{We measure a total speedup of 3 compared to the naive implementation in test simulations on a Intel Haswell Core i7-5600U 2.6 GHZ CPU with first order DG base functions on a structured $32 \times 32 \times 32$ mesh for subsurface and $32 \times 32$ grid for surface flow. Even higher speedups are expected with higher-order base functions and matrix-free iterative solvers.
The fully-coupled combination of the Richards solver obtained with the code generation framework and surface-runoff is tested with DG up to fourth order on structured as well as on unstructured grids. Parallel simulations are possible as well.
}{We measure a total speedup of 3 compared to the naive implementation in test simulations on a Intel Haswell Core i7-5600U 2.6 GHZ CPU with first order DG base functions on a structured $32 \times 32 \times 32$ mesh for subsurface and $32 \times 32$ grid for surface flow. Even higher speedups are expected with higher-order base functions.
The fully-coupled combination of the Richards solver obtained with the code generation framework and surface-runoff is tested with DG up to third order on structured and first order on unstructured grids. Parallel simulations are possible as well. It was already used to simulate larger regions with topographical data taken from a digital elevation model (Figure~\ref{pic:landscape}). Large-scale scalability and performance tests are currently in preparation and will be conducted on the HLRN-IV cluster in Göttingen and on SuperMUC-NG in Munich.}

\section{Conclusion}

In \exadune{} we extended the DUNE software framework in several directions to make it ready for the
exa-scale architectures of the future which will exhibit a significant increase in node level performance
through massive parallelism in form of cores and vector instructions.
Software abstractions are now available that enable asynchronicity as well as parallel exception
handling and several use cases for these abstractions have been demonstrated in this paper: resilience in multigrid
solvers and several variants of asynchronous Krylov methods. Significant progress has been achieved in hardware-aware
iterative linear solvers: we developed preconditioners for the GPU based on approximate sparse inverses, developed matrix-free
operator application and preconditioners for higher-order DG methods
and our solvers are now able to vectorize over multiple right hand sides. These building blocks have then been used to implement
adaptive localized reduced basis methods, multilevel Monte-Carlo methods and a coupled surface-subsurface flow solver on up to 14k cores.
The \exadune{} project has spawned a multitude of research projects, running and planned, as well as further collaborations
in each of the participating groups. We conclude that the \dune{} framework has made a major leap forward due to the
\exadune{} project and work on the methods investigated here will continue in future research projects.

\bibliographystyle{plain}
\bibliography{exadune}

\end{document}